\documentclass[fleqn,usenatbib]{mnras}

\usepackage{newtxtext,newtxmath}
\usepackage{hyperref}

\hypersetup{colorlinks=true, allcolors=blue}

\usepackage[T1]{fontenc}

\DeclareRobustCommand{\VAN}[3]{#2}
\let\VANthebibliography\thebibliography
\def\thebibliography{\DeclareRobustCommand{\VAN}[3]{##3}\VANthebibliography}

\usepackage{graphicx}	
\usepackage{amsmath}	
\usepackage{color}

\title[Transients in LoTSS - framework development]{Transient study using LoTSS - framework development and preliminary results}

\author[I. de Ruiter et al.]{
Iris de Ruiter$^{1}$\thanks{E-mail: i.deruiter@uva.nl},
Zachary S. Meyers $^{2,3}$, 
Antonia Rowlinson$^{1,4}$,
Timothy W. Shimwell$^{4,5}$,
David Ruhe$^{1,6}$,
\newauthor
Ralph A.M.J. Wijers$^{1}$
\\
$^{1}$Anton Pannekoek Institute for Astronomy, University of Amsterdam, Science Park 904, NL-1098 XH Amsterdam, the Netherlands\\
$^{2}$Deutsches Elektronen-Synchrotron DESY, Platanenallee 6, 15738 Zeuthen, Germany\\
$^{3}$Erlangen Center for Astroparticle Physics (ECAP), Friedrich-Alexander-Universität Erlangen-Nuremberg, 91058 Erlangen, Germany\\
$^{4}$ASTRON, the Netherlands Institute for Radio Astronomy, Postbus 2, NL-7990 AA Dwingeloo, the Netherlands\\
$^{5}$Leiden Observatory, Leiden University, P.O. Box 9513, NL-2300 RA Leiden, the Netherlands\\
$^{6}$AI4Science (AMLab), Informatics Institute, University of Amsterdam, The Netherlands
}

\date{Accepted XXX. Received YYY; in original form ZZZ}

\pubyear{2023}

\begin{document}
\label{firstpage}
\pagerange{\pageref{firstpage}--\pageref{lastpage}}
\maketitle

\begin{abstract}
We present a search for transient radio sources on time-scales of seconds to hours at 144 MHz using the LOFAR Two-metre Sky Survey (LoTSS). This search is conducted by examining short time-scale images derived from the LoTSS data. To allow imaging of LoTSS on short time-scales, a novel imaging and filtering strategy is introduced. This includes sky model source subtraction, no cleaning or primary beam correction, a simple source finder, fast filtering schemes and source catalogue matching. This new strategy is first tested by injecting simulated transients, with a range of flux densities and durations, into the data. We find the limiting sensitivity to be 113 and 6 mJy for 8 second and 1 hour transients respectively.
 The new imaging and filtering strategies are applied to 58 fields of the LoTSS survey, corresponding to LoTSS-DR1 (2\% of the survey). One transient source is identified in the 8 second and 2 minute snapshot images. The source shows one minute duration flare in the 8 hour observation. Our method puts the most sensitive constraints on/estimates of the transient surface density at low frequencies at time-scales of seconds to hours; $<4.0\cdot 10^{-4} \; \text{deg}^{-2}$ at 1 hour at a sensitivity of 6.3 mJy; $5.7\cdot 10^{-7} \; \text{deg}^{-2}$ at 2 minutes at a sensitivity of 30 mJy; and $3.6\cdot 10^{-8} \; \text{deg}^{-2}$ at 8 seconds at a sensitivity of 113 mJy. In the future, we plan to apply the strategies presented in this paper to all LoTSS data.
\end{abstract}




\section{Introduction}
The transient radio sky provides a unique opportunity to study the most extreme events that take place in the Universe. The astrophysical processes that generate transient phenomena are often highly dynamic and explosive, allowing us to study environments that are inaccessible on Earth. Transient or highly variable sources have been observed across all wavelengths, however, radio astronomy offers a different perspective as some astrophysical phenomena are either highly beamed, unique to radio frequencies or obscured by dust at other wavelengths. Over the last two decades, radio transients have been discovered all across the transient phase space, which spans orders of magnitude in transient time-scales, observing frequency and flux density. There are several astrophysical phenomena that are known to be transient at low radio frequencies. These include events like stellar flares, magnetar flares, novae, X-ray binaries, intermittent pulsars, FRBs and strongly scintillating AGN. In this study, we focus on searching for low-frequency (144 MHz) radio transients with durations of seconds to hours. To this end, we use survey data obtained with the Low Frequency Array (LOFAR; \cite{van2013lofar}). Our search is sensitive to various transient phenomena. The most relevant source classes for this study are giant pulses from pulsars and flare stars \citep{spangler1976simultaneous, callingham2021low, feeney2021detection},
coronal mass ejections \citep{crosley2018low}, 
X-ray binaries \citep{chandra2017giant, chauhan2021broadband, monageng2021radio}, and possibly Algol-type binaries \citep{lefevre1994variability, umana1998radio}. Recently, long-period magnetars are confirmed to exist and these are also detectable in the low-frequency image plane \citep{hurley2022radio, hurley2023long, caleb2022discovery}. The references here point specifically to low-frequency radio detections of these phenomena within the aforementioned time-scales. We note that Algol-type binaries have mainly been studied at higher radio frequencies (>1 GHz) and the variability time-scale of X-ray binaries might be too long ($\sim$ days) to probe with this study. Finally, strongly scintillating background AGN could be interpreted as variables or even transient sources. An overview of radio transient phenomena at various time-scales can be found in Figures 3 and 5 in \cite{pietka2015variability}.

Additionally, there are several theories predicting low-frequency, coherent radio emission from short gamma-ray bursts from compact binary mergers, see e.g. \cite{rowlinson2019constraining} and references therein. Some of these models suggest that the emission mechanisms of these types of sources is similar to fast radio bursts (FRBs), which are another target of this study due to the low observing frequency of LOFAR. Image domain searches for FRBs \citep{tingay2015search, rowlinson2016limits, andrianjafy2023image, driessen2023frb} utilize the signal delay introduced by the dispersion measure. The signal delay $ \Delta t \approx 4.15 (\nu_{\mathrm{lo}}^{-2} - \nu_{\mathrm{hi}}^{-2}) \rm{DM} \; \rm{ms}$ is defined in terms of the lower and upper limit of the observation bandwidth in GHz, $\nu_{\mathrm{lo}}$ and $\nu_{\mathrm{hi}}$ respectively, and the dispersion measure, DM (See Eqn. 1 in \cite{petroff2019fast}). For the LoTSS bandwidth of $0.120$ to $0.168$ GHz \citep{shimwell2019lofar} and a typical DM of 500 $\rm{cm}^{-3} pc$ (see Figure 6 in \cite{chawla2022modeling} and Figure 1 in \cite{arcus2021fast}) one thus expects a signal delay of $70.6 \cdot 10^3$ ms. The FRB signal will be spread out over 70 seconds, implying that a bright burst could be detected in the image domain.

Furthermore, there is a possibility to probe new source classes. Examples of these are frequent in the low-frequency radio sky. A famous example is the class of Galactic centre Radio Transients \citep{hyman2002low, hyman2005powerful, hyman2009gcrt}, whose bursts last from minutes to months. Specifically, a source like the `Galactic Burper', which has shown a series of $\sim 1$ Jy bursts \citep{hyman2005powerful} and single bursts years later \citep{hyman2006new, hyman2007faint}, would be an ideal target for our study. \cite{jaeger2012discovery} find a low-frequency radio transient that is variable on a time-scale of hours. The source has no counterpart which makes identification difficult, but several characteristics point towards a stellar flare. \cite{obenberger2014limits} find two extremely bright transients, at peak flux densities of $\sim 3$ kJy, lasting for $\sim 100$ seconds at 30 MHz. Finally, \cite{stewart2016lofar} find a bright (possibly Galactic) transient towards the North Celestial Pole at 60 MHz, lasting for <10 min. These transients discussed in this paragraph are either difficult to ascribe to any of the known source classes, or could possibly be detections of exciting new source classes.\\

Traditionally, most detections of coherent emitters have been done with time-series techniques (time-domain), while incoherent emission, which generally evolves over longer time-scales, is observed in the image plane (image-domain). Image-domain studies make snapshot images of a patch of sky and use those to look at the time variability of sources. For coherent emission processes, the accelerated particles can cooperate in phase, resulting in emission that can reach extremely high brightness temperatures. Incoherent emission comes from the summation of the radiation from individual accelerated particles and therefore the brightness temperature is limited to $\sim 10^{12}$ K. The stellar flare mechanisms are usually coherent emission mechanisms at 144 MHz, as well as the emission from magnetars (related to short gamma-ray bursts or FRBs), while most of the other source classes operate through incoherent emission mechanisms, usually synchrotron emission. This work presents an image domain transient study that probes both coherent and incoherent processes. \\

In order to quantify the number of transients we expect to see in the low-frequency radio sky, we determine the rate of transient sources at various time-scales and flux densities. To obtain the most constraining value of the transient surface density, observations of large sky areas at high sensitivities are required. This paper performs a transient search in the image domain using the LOFAR Two-metre Sky Survey \citep[LoTSS]{shimwell2019lofar, shimwell2022lofar}. LoTSS images the low-frequency northern sky at unprecedented sensitivity and resolution.  Section \ref{sec:methods} presents a method to create snapshot images at 8 second, 2 minute and 1 hour time-scales from the 8 hour LoTSS pointings, and search for transients in the snapshot images. Section \ref{sec:sensitivity} discussed the sensitivity limits of this method via simulated transient sources. We apply this method to a preliminary data set of 58 LoTSS-DR2 pointings (corresponding to the coverage of LoTSS-DR1) and present the results in Section \ref{sec:results}. The implications of these results are discussed in Section \ref{sec:discussion} and finally, our concluding remarks and outlook for future application of our methods, are presented in Section \ref{sec:conclusion}.

\section{Methods} \label{sec:methods}
\begin{figure*}
    \includegraphics[width=1.1\textwidth]{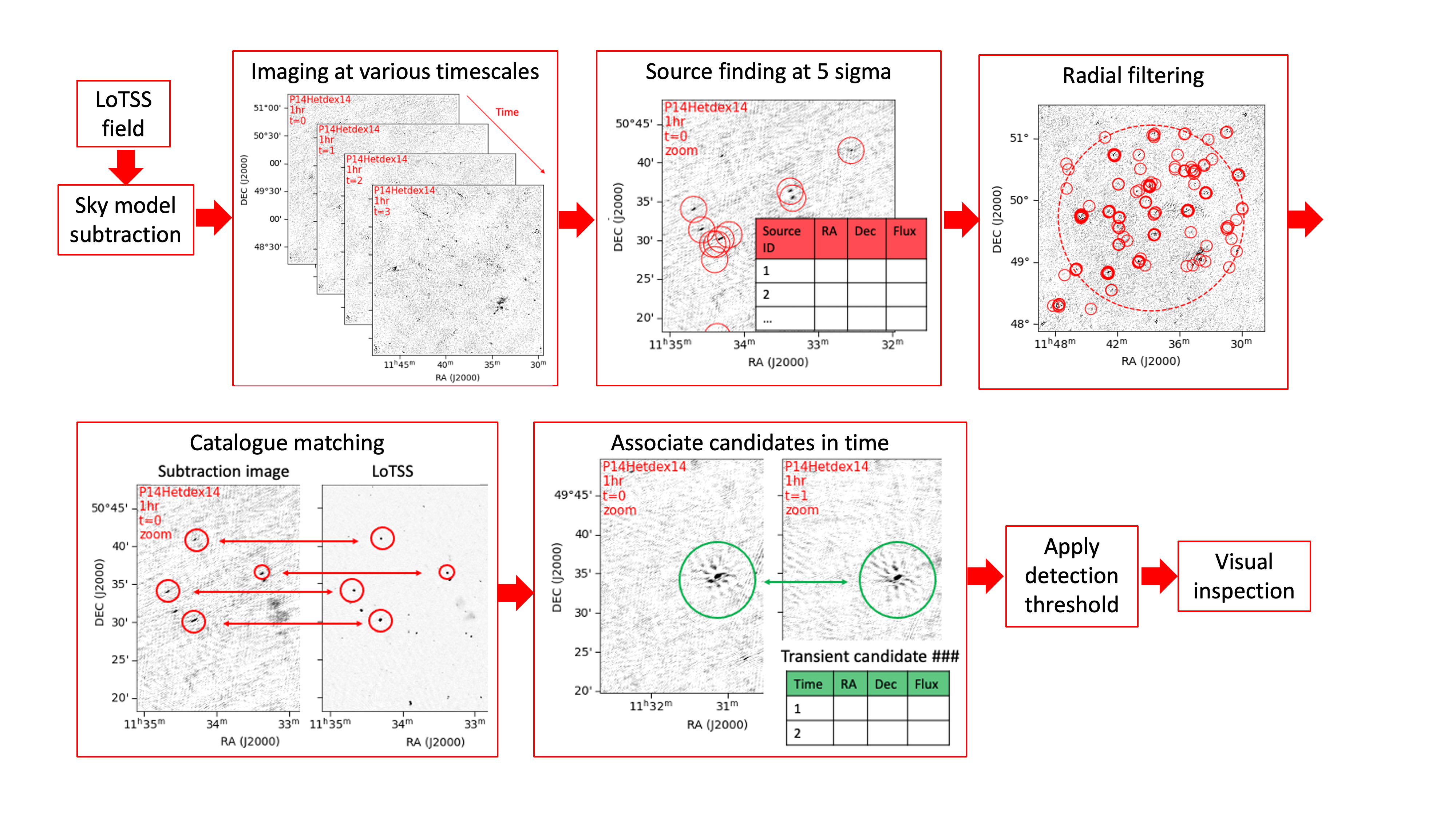}
    \caption{Illustration of our method to search for transients on snapshot time-scales of 8 seconds, 2 minutes and 1 hour in each 8 hour LoTSS pointing. Each box corresponds to a subsection in the Methods.}
    \label{fig:process}
\end{figure*}

The next subsections will discuss each step in our method to search for transients on snapshot time-scales of 8 seconds, 2 minutes and 1 hour in each 8 hour LoTSS pointing. Figure \ref{fig:process} shows a visual schematic of the process. Each box in Figure \ref{fig:process} corresponds to one subsection. The method presented here is meant to identify transient candidates that leave no signature in the deep 8-hour integration image. Variability of sources in the deep image is currently excluded from this analysis. The method described below is developed to identify transient candidates in an efficient manner, follow-up with more traditional and elaborate imaging approaches is then necessary to fully characterize the transient source.

\subsection{LOFAR Two-Metre Sky Survey}
LOFAR \citep{van2013lofar} is a low-frequency radio telescope that is comprised of many thousands of dipole antennas arranged in stations. These stations are distributed all over the Netherlands and more sparsely throughout Europe. The LOFAR Two-Metre Sky Survey (LoTSS; \cite{shimwell2017lofar}) aims to image the whole northern sky in 3168 pointings. The survey has had two major data releases so far, DR1 \citep{shimwell2019lofar} covering 58 pointings and DR2 \citep{shimwell2022lofar} covering 814 pointings. LoTSS observes between 120 and 168 MHz. The flux densities are given at the central frequency of 144 MHz. 

Whilst LoTSS data contains the entire international baseline coverage of LOFAR the data releases to-date only contain the Dutch stations, yielding a maximum baseline of 121 km \citep{van2013lofar} resulting in an image resolution of $6 \arcsec$, due to computational limitations. This resolution combined with a median RMS of 83 $\mu \rm{Jy} \; \rm{beam}^{-1}$ for the low-frequency continuum images allows LoTSS to venture into a realm of the radio sky that has been unexplored up to now.\\

In this work, we perform our transient study on the 58 pointings that cover the first data release of LoTSS \citep{shimwell2019lofar}, but we note that we do use LoTSS-DR2 products that have been processed according to \cite{shimwell2022lofar}. To date, the LoTSS data releases have only included regions at high Galactic latitudes. However, the goal of LoTSS is to image the whole Northern sky, including the Galactic Plane \citep{shimwell2022lofar}. The Galactic Plane is the most promising region for most transient candidates discussed in the introduction. One of the goals of this paper is to establish a framework that can be applied to future LoTSS data releases, including the Galactic Plane.

\subsection{Sky model subtraction} \label{sec:methods_sky_model_subtraction}

A technique that is often employed in transient studies is the use of \textit{difference imaging}; once you create all snapshot images, subtract images of consecutive timesteps, resulting in a difference that should allow for easier transient identification. However, this technique is not well suited for radio transient studies, because with many facilities it is challenging to create good-quality images for each timestep due to sparse uv-coverage giving an irregular point spread function (PSF). Due to the irregular PSF and structured noise in individual images, a large number of artefacts are created when subtracting subsequent images. Therefore, the difference imaging technique is not well-suited for radio transient searches.  

As an alternative, here we use source \textit{subtracted data} to search for transients (see eg. \cite{fijma2023new, wang2023radio}).  Source subtracted data are created by subtracting the full-observation sky model from the uv-plane using DDFacet \citep{tasse2018faceting, tasse2021lofar} to apply the direction-dependent calibration solutions during the subtraction. The images at various time-scales created from these data show the difference between the sky during the snapshot time and the full observation sky model. A more mathematical description of this technique can be found in Appendix \ref{app:maths}. The source subtracted data is imaged at various cadences and because most of the signals are removed by subtracting out the sky model, the adverse effects of the poor uv-coverage and associated PSF in the images are limited.



There are three major benefits to working with source subtracted data, compared to traditional methods. First of all, subtraction of the sky model from the uv-plane greatly reduces the compute time spent on imaging, because no primary beam correction or cleaning is required, as there are typically no sources in the field. A reduction of the imaging time is critical when imaging a full 8-hour observation on an 8 second cadence. An additional benefit is that we can create source subtracted images even when the PSF is poorer due to sparse uv-sampling. Secondly, the subtracted images should, in theory, only contain the sources that are not in the sky model, or sources with a variable flux density compared to the sky model, which simplifies a transient search. Lastly, by subtracting the full 8-hour sky model from the shorter time-scale snapshots, one can remove the high confusion noise from the snapshots, which allows for a deeper transient search (see eg. \cite{fijma2023new}). This is because the removal of the sources allows one to image at lower resolution without being limited by confusion noise. \\

One aspect that we want to point out is that the direction-dependent calibration solutions have time-variable behaviour. Therefore, a potential problem of this method is that the solutions could absorb an actual transient in order to make the result look like the sky model against which the data is calibrated. To this end, we perform simulations where we inject a transient source before calibration (against a sky model without the transient source) and analyze the result. See Section \ref{sec:simulations} for more details on these simulations.

In conclusion, the subtraction imaging method should be able to probe any variable and transient behaviour on time-scales shorter than the duration of the full sky model observation, which is 8 hours in the case of LoTSS. A script\footnote{\url{https://github.com/mhardcastle/ddf-pipeline/blob/master/scripts/sub-sources-outside-region.py}} to perform the sky model subtraction exists within the pipeline used for LoTSS processing (DDF-pipeline \cite{tasse2018faceting, tasse2021lofar})\footnote{\url{https://github.com/mhardcastle/ddf-pipeline}} and yields a single sky model subtracted measurement set for each LoTSS field (or pointing). Throughout the rest of this work, when referring to \textit{subtracted images}, we refer to the sky model subtracted images.\\

\subsection{Snapshot imaging} \label{sec:methods_snapshot_imaging}

After the sky model has been subtracted, images are produced from the subtracted data column in the measurement sets using {\sc WSClean} \citep{offringa2014wsclean}. The subtracted data is imaged on cadences of 8 seconds, 2 minutes and 1 hour, which yields a large number of images. For example, the 8 second cadence will yield 3600 images for the full 8 hour observation per field. Therefore, imaging parameters are chosen to minimize the compute time spent on imaging. No CLEAN algorithm iterations need to be applied because our first goal is to find any emission that is not subtracted out (ie. transients sources). To this end, there is no need to deconvolve the PSF from the dirty image.
Additionally, no primary beam correction is performed because we do not need accurate flux densities to identify transient sources. Furthermore, transient sources should be easier to identify against a more uniform noise background. An example of a subtracted snapshot image is given in Figure \ref{fig:sub_image_art_example}. The left panel shows an 8 second integration subtracted image and the right panel shows the corresponding part of the sky in the full integration 8 hour LoTSS data. Most sources have been subtracted out nicely, except for the bright 2.1 Jy sources in the centre of the image, which shows an artefact of inaccurate source subtraction in the subtracted image. Section \ref{sec:methods_source_catalogue_matching} elaborates on how we mitigate these particular artefacts.

The most important {\sc WSClean} parameters include a pixel size of 6", implying that the LOFAR Dutch station baseline resolution is mapped to one pixel. The LoTSS pointings are typically separated by 2.58\textdegree 
 \citep{shimwell2019lofar}, and to ensure substantial overlap we image $3.67$x$3.67^\circ$ for each pointing. Finally, an important parameter is the padding factor, which specifies the factor by which the image size is increased beyond the field of interest to avoid edge issues. We found it was crucial to increase this parameter from its default value of $1.2$ to $1.6$ to reduce some of the fake transient source introduced by aliasing effects. Details on this issue are given in Appendix \ref{app:psf_artefacts}. The {\sc WSClean} imaging parameters are summarised in Table \ref{tab:wsclean}. 

\begin{table}
\begin{tabular}{ll} \hline
Setting                    & Value        \\ \hline
Number of clean iterations & 0            \\
Size of image (pixels)     & 2200         \\
Size of one pixel (arcsec) & 6            \\
(Min, max) uv-l             & (50,60000) \\
Briggs weighting           & -0.25       \\ 
Intervals out              & 8, 225, 3600 \\
Padding                    & 1.6 \\ \hline
\end{tabular}
\caption{\label{tab:wsclean} The {\sc WSClean} settings used to image all the observations presented in this analysis. All other settings are default settings.}
\end{table}


\subsection{Source finding} \label{sec:methods_source_finding}

We use the `Live Pulse Finder' software presented in \cite{lpf2021soft, ruhe2022detecting} to perform source finding in the subtracted images. 
Using spatial convolutions, the \textsc{LPF} source finder creates a detailed model of the noise in each image, which is necessary since the subtracted images show noise structures that may not be accurately captured by traditional source finding methods. After constructing a noise model the source finder looks for peaks in signal-to-noise (SNR) ratio above the prespecified threshold and saves the locations. Since this source finder simply identifies pixels that significantly stand out from the noise it is fast compared to methods that perform Gaussian fits to the source shape (eg. \textsc{PySE} \cite{carbone2018pyse}).

We perform a blind search for sources in our subtracted images with a detection threshold of $5\sigma$. This detection threshold is rather low for the enormous number of pixels that is searched. In Section \ref{sec:methods_detection_thresholds} this value is updated to a more stringent detection threshold. We decide to perform the source finding for $5\sigma$ sources as this will allow us to characterize transient candidates in more detail later on. For example, a marginal detection above the stricter detection threshold in one snapshot could be accompanied by several subthreshold detections in adjacent snapshots at the same location. This information helps to decide whether or not a transient candidate is a potential imaging artefact (see Section \ref{sec:methods_detection_thresholds}).




\subsection{Radial filtering} \label{sec:methods_radial_filtering}
After finding sources in the full $3.67$x$3.67^\circ$ square subtracted image, a radial filter is applied. As the image quality decreases with radial distance to the centre of the pointing \cite{shimwell2022lofar}, we filter out all sources that lie more than 1.5$^{\circ}$ away from the beam/pointing centre. This still ensures significant overlap between pointings (See Section \ref{sec:methods_snapshot_imaging}) but allows us to disregard noisier parts of the image.

\subsection{Source catalogue matching} \label{sec:methods_source_catalogue_matching}
Ideally, the subtracted image only shows sources that have a significantly different flux density at the time step considered, compared to the full image. In reality, the subtracted image also contains artefacts that arise mainly around bright and/or extended constant sources.

\begin{figure}
    \centering
    \includegraphics[width=\linewidth]{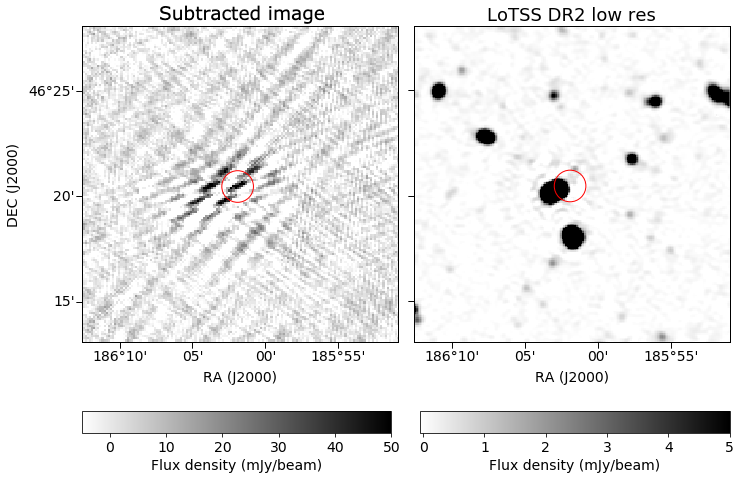}
    \caption{Imaging artefact around a 2.1 Jy source in an 8 second subtracted image snapshot. The left panels shows the subtracted image, while the right panels shows the corresponding sky area in the LoTSS survey.}
    \label{fig:sub_image_art_example}
\end{figure}

These subtraction artefacts are not the transient candidates that are targeted in this study and should be filtered out. Figure \ref{fig:sub_image_art_example} shows the importance of this particular step in the analysis. The left panel shows an 8 second snapshot subtracted image and the right panel shows the LoTSS image of the corresponding part of the sky. The red circle indicates the location of a source identified by the source finder. In fact, all separate parts of the artefact in the subtracted image are identified as separate sources by the source finder. However, looking at the LoTSS image (right panel Figure \ref{fig:sub_image_art_example}) it is clear that these sources are a result of improper subtraction of the central, bright 2.1 Jy source. We therefore want to disregard all these sources as transient candidates, which is done via the catalogue matching scheme detailed below.

\begin{figure*}
    \centering
    \includegraphics[width=\textwidth]{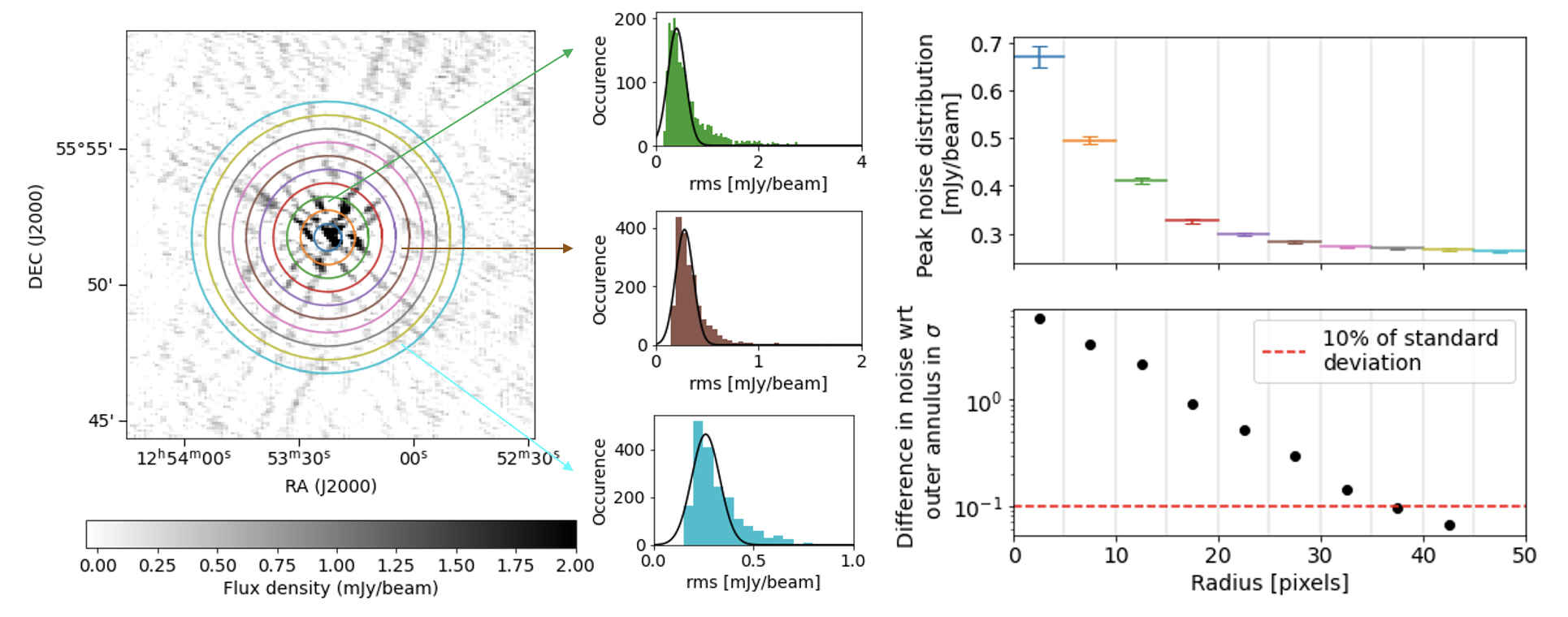}
    \caption{Example of determining a filter radius around artefacts of known sources with flux 100-500 mJy in 1 hour subtract snapshots. The left panel shows an example 1 hour subtracted image with a significant structure due to improper subtraction of a 314 mJy source at that location. We define different annuli around this source and investigate the noise properties; examples of the pixel distributions in three of these annuli are shown in the plots in the middle of this figure. The top right panel shows the noise in each annulus for a sample of artefacts around 100-500 mJy in 1 hour subtract snapshots. The bottom right panel shows the difference between the peak of the noise distribution in each respective annulus compared to the outer annulus. This difference is shown in standard deviations of the noise distribution of the outer annulus. The filter radius is defined at the point where the noise distribution starts to deviate by more than 10\% of the standard deviation of the outer noise distribution, indicated by the red dashed line. }
    \label{fig:artefact_extent_investigation}
\end{figure*}

In order to define regions around known sources that are disregarded in the transient search, we first investigate up to which radius to filter around known sources in the subtracted image. We perform this investigation separately on the three snapshot time-scales used in this study. To this end, we gather known sources in the following integrated flux density intervals [10, 50], [50,100], [100,500], [500,1000] and >1000 mJy and define a 'filter' radius for each of these flux categories. An example of this investigation for 1 hour subtracted snapshots targeting 100-500 mJy known sources is shown in Figure \ref{fig:artefact_extent_investigation}. The left panel shows an example of a 1 hour subtracted image. The image shows significant structure, which is due to a known 314 mJy source at this location. Different coloured annuli around this structure define different trial filter radii. The histograms in the middle show examples of pixel distributions in these annuli, where the noise goes down for the annuli with larger radii. The top plot in the right panel shows the mode of the pixel histogram distribution in each annulus. This represents the noise level in each annulus. We now assume that the noise distribution in the outer annulus is representative of the local noise distribution and compare the noise distribution in each annulus to the outer annulus. The bottom panel shows the difference between the peak of the noise distribution for each respective annulus compared to the outer annulus, defined in terms of the standard deviation of noise distribution in the outer annulus. Once the noise difference becomes more than 10\% of the standard deviation of the noise distribution in the outer annulus, we set a filtering limit. This is the point where the subtraction artefacts start to significantly impact the noise distribution. The red dashed line in the bottom right panel shows this cutoff. In this case setting a filter radius of 35 pixels, corresponding to $3.5\arcmin$. Note that although the left panel shows a single instance of an artefact around a  100-500 mJy known source in a single 1 hour snapshot, the histograms and plots on the right side are created from a representative sample of such sources.

The process outlined in this section is repeated for each of the flux density intervals and each snapshot time-scale. This results in the filtering radii summarized in Table \ref{tab:filtering_radii}. Catalogue sources with flux density between 10-50 mJy did not give significant improper subtraction artefacts in the 2 minute and 8 second subtracted snapshots, therefore no filtering is necessary there. The same holds 50-100 mJy source in the 8 second subtracted snapshots. In summary, transient candidates that lie within the filter radius of catalogue sources (as defined in Table \ref{tab:filtering_radii}), are disregarded in the further steps of the transient search.

\begin{table}
\begin{tabular}{ll|lll}
Lower limit  & Upper limit & \multicolumn{3}{|l|}{Filter radius [arcmin]}  \\
flux [mJy] & flux [mJy] & 1 hr & 2min & 8sec \\ \hline
1000     & $\infty$        & 4  & 4  & 3    \\
500      & 1000            & 3.5  & 3  & 2.5      \\
100      & 500             & 3.5  & 2    & 1.5      \\
50       & 100             & 2  & 1.5    & -      \\
10       & 50              & 1    & -    & -     
\end{tabular}
\caption{Filter radii around catalogue sources determined for various flux density intervals and the three imaging time-scales considered in this study. The filter radii are in arcmin around the location of the source in the LoTSS catalogue. \label{tab:filtering_radii}}
\end{table}

\subsection{Associate candidates in time}\label{sec:methods_associated_candidates_in_time}
All steps detailed above were performed on individual snapshots. The next step is to group together transient candidates that are found in multiple images of the same snapshot time-scale. For example, if the first 100 images of the 8 second snapshots contain a transient candidate at roughly the same location (within $5\arcmin$), these candidates are grouped together as one individual candidate. This way this source only has to be visually inspected in a few of these images before deciding whether it is an artefact or an actual interesting transient candidate.

\subsection{Detection threshold} \label{sec:methods_detection_thresholds}
So far we have not done any filtering on the detection significance of transient candidates. The source finder (Section \ref{sec:methods_source_finding}) identifies all sources with a signal-to-noise ratio $\geq 5$. This value is low and will yield false positive transient identifications because a large number of pixels is trialed. A detection threshold is calculated based on the probability that one pixel is encountered that exceeds the detection threshold for the total number of pixels per imaging time-scale. In other words, we calculate the probability $P(X\leq x) = 1 - \frac{1}{N}$ with N the total of number of pixels sampled per time-scale.

We calculate the value of $x$ expressed in $\sigma$ (i.e. the threshold) for this probability using the percent point function (or inverse cumulative distribution function) for the Gaussian distributions of the images per time-scale using the scipy stats norm ppf functionality \citep{virtanen2020scipy}. The results of this calculation are shown in Table \ref{tab:detection_thresholds}. Note that for we perform this calculation for a random subset of all images, but repeating this procedure for a different subset gives similar results implying that the subset is representative of the full sample. Finally, the areas around known catalogue sources are not searched in for transients but those cuts are not taken into account here. The full inner region of the images ($\pi*900^2=2544690$ pixels per image) is assumed to be searched in this calculation. Accounting for the cut areas does not make much difference, for the 1 hour subtracted images up to about 20\% of the image is cut out, but this results in the detection threshold being lowered to $5.99\sigma$ (as opposed to $6.02\sigma$, see Table \ref{tab:detection_thresholds}).

\begin{table}
\begin{tabular}{lll}
Time scale & Number of pixels (N) & Detection threshold \\ \hline
1 hr       & $1.2\cdot 10^9$ & $6.02 \sigma$       \\
2 min      & $3.3\cdot 10^{10}$ & $6.54 \sigma$       \\
8 sec      & $5.3 \cdot 10^{11}$ & 6.94 $\sigma$      
\end{tabular}
\caption{Detection thresholds for transient candidates to be considered for visual inspection. \label{tab:detection_thresholds}}
\end{table}

We apply these thresholds to all light curves that come out of the previous step. A light curve is put forward for visual inspection if at least one source in the time series has a signal-to-noise ratio above the threshold calculated here. This is done such that other subthreshold detections of the same transient candidate can be considered in visual inspection.

\subsection{Visual inspection}

After the filtering process, there are $\mathcal{O}(10)$ transient candidates left per field for all time-scales. There are fields, for example \textsc{P173+55}, that have one very bright source (3.7 Jy) that affects the sky model subtraction over a large fraction of the field and produces many spurious candidates. The remaining sources are visually inspected. There are two main categories of false positive transient candidates that still come through the pipeline. First of all, there are subtraction artefacts due to faint sources that generally do not give significant subtraction artefacts. Secondly, there are bright known sources that show extended subtraction artefacts, that are not filtered out by the filter radii presented in Table \ref{tab:filtering_radii}. Additional filtering, around fainter sources or extending the filter radii around bright sources, would solve both issues but comes at the cost of a further reduced sky area that is searched for transients. The final visual inspection step thus removes transient candidates that are associated with faint sources in the deep field, for which we have not applied any filtering radius, or transient candidates that are associated with artefacts of bright sources that extend beyond the filtering radius.

\section{Sensitivity} \label{sec:sensitivity}
\subsection{Simulations} \label{sec:simulations}

\begin{figure}
    \includegraphics[width=\linewidth]{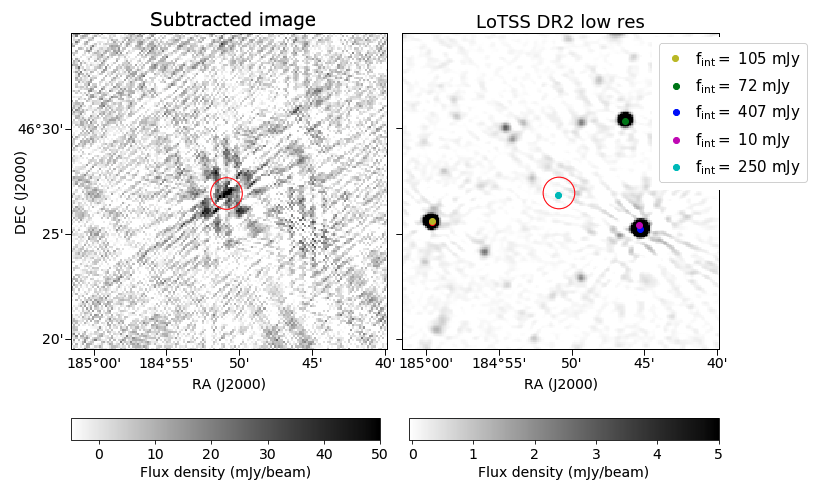}
    \caption{Simulated transients source of 250 mJy with a duration of 8 seconds, captured by one 8 second subtracted image snapshot. The left panel shows the subtracted image, while the right panel shows the corresponding sky area in the LoTSS survey. The colored dots and legend indicate the flux densities of the sources in the LoTSS source catalogue.}
    \label{fig:8sec_sim}
\end{figure}

To assess the sensitivity of our method we perform simulations of transient sources. To this end, transient sources are injected into a field, and the data is recalibrated. Here, the full direction-dependent calibration is repeated to investigate whether or not the calibration strategy might absorb faint short-duration transients. Afterwards the subtraction imaging and filtering methods described in Section \ref{sec:methods} are applied. By injecting simulated transient sources with different integrated flux densities throughout the field, the sensitivity of the method as a function of distance from the beam centre can be determined. For these simulations, the P23Hetdex20 field is used, which is a typical field containing a variety of diffuse, moderately bright sources and normal levels of calibration artefacts. 

8 second transients are injected with integrated flux densities of [70.0, 80.2, 96.5, 113.3, 133.0] mJy. 1 hour transients are injected with integrated flux densities of [0.2, 0.6, 2.0, 6.3, 20.0] mJy. Each flux density value is injected at 5 different radii across the whole field, because we expect a dependence of sensitivity as a function of radius, where the search is less sensitive away from the field centre, due to the primary beam. We inject simulated transients such that they do not overlap with bright LoTSS sources, as those would be filtered out via the steps described in Section \ref{sec:methods_source_catalogue_matching}. 

Figure \ref{fig:8sec_sim} shows an example of a simulated transient source. The left panel shows the injected transient in the subtracted image. On the right, the LoTSS image is shown for this part of the sky for comparison. The simulated transient is injected into a relatively empty part of the sky. Furthermore, the subtracted image shows that in this case the subtraction has worked quite well and the injected transient dominates everything else in the nearby region. This particular simulated transient has a duration of 8 seconds and an integrated flux density of 250 mJy.  \\
\begin{figure}
    \centering
    \includegraphics[width=\linewidth]{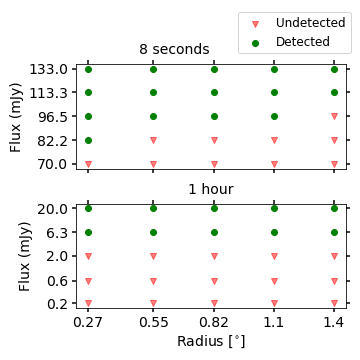}
    \caption{Results of the injecting simulated transients with various flux densities (y-axis) at different radii throughout the field (x-axis). The top panel shows the 8 second duration simulated transients, the bottom panel shows the 1 hour duration simulated transients.}
    \label{fig:sim_results}
\end{figure}

We inject the transient sources in the data in such a way that they will be captured by exactly one snapshot, ie. the transient is injected at a time where it will fully fall within one snapshot image and is not captured partly by two consecutive snapshot images. After injecting the transients in the field, we recalibrate the data against the LoTSS sky model that does not contain the transient. This is to check whether or not a transient can be absorbed into the calibration solutions, and therefore might appear differently than initially expected. The recalibrated data are then processed via the exact same steps outlined in Figure \ref{fig:process} and Section \ref{sec:methods}. Figure \ref{fig:sim_results} shows which of the injected simulated transients were recovered by our method. The plot shows the flux densities of the injected transients injected at different radii, where the green dots indicate that the particular transient was recovered at the end of the pipeline, while the red triangles indicate that the transient was not found by our pipeline. The sensitivity is fairly constant throughout the beam for this set of simulations for the 1 hour transients. For the 8 second transients a relation between the flux of the recovered simulation and radius is visible. We draw the conservative conclusion that the sensitivity of the 1 hour transient search is around 6 mJy and around 113 mJy for the 8 second transient search. Extrapolating the sensitivity of 6.3 mJy at 1 hour and 113 mJy at 8 seconds to 2 minutes, using that the detection sensitivity scales with $~1/\sqrt{t}$, we find a sensitivity of roughly 30 mJy for the 2 minute snapshots.

We want to note that these sensitivity estimates are lower limits because we assume that the transient is exactly covered by one snapshot. In reality, an 8 second transient will not exactly line up with the binning of our snapshots, and therefore we will not reach the aforementioned sensitivity of 113 mJy, as the source is split up over two snapshots. The worst case limits for detections is when the fluence of a transient is split equally between two bins, in which case the limits should be multiplied by a factor $\sqrt{2}$, which yields detection thresholds of [160, 42, 8.9] for the 8 second, 2 minute and 1 hour snapshots respectively.

\subsection{Upper limits}
\begin{figure}
    \centering
    \includegraphics[width=\linewidth]{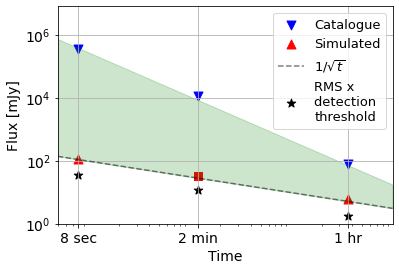}
    \caption{The green shaded area shows the flux density of potential transient sources as a function of time-scale that the transient search method presented in this work is sensitive to. The upper limits indicated with blue triangles are based on the filtering of known catalogue sources outlined in \ref{sec:methods_source_catalogue_matching} and Table \ref{tab:filtering_radii}. The lower limits are determined using simulations, as outlined in Section \ref{sec:simulations}. The black crosses are the lower limits of our search as determined by multiplying the rms noise in the subtracted images (Section \ref{sec:results_image_quality}) by the detection thresholds (Section \ref{sec:methods_detection_thresholds}). }
    \label{fig:sensitivity}
\end{figure}

The previous section determines the lower sensitivity limit to the transient search. In this section the upper limit is explored. When transient sources get extremely bright, they might get into the sky model and show up on the deep image, despite being active only a fraction of the time of the full observation. As explained in Section \ref{sec:methods_source_catalogue_matching} the sources in the deep image are used to filter out subtraction artefacts in the subtracted images, but this implies that also these extremely bright transients will be filtered out. A simple estimate of the brightest source this method would find per timestep is given by multiplying the lowest flux density value included in our filtering scheme as outlined in Table \ref{tab:filtering_radii} by the number of snapshots that are made for a particular time-scale. For example, for the 1 hour snapshots filters are applied around deep field sources of 10 mJy. A transient source that is active for 1 hour would therefore have to emit at $10\cdot 8=80$ mJy to make it into the deep image and be disregarded in the transient search. Figure \ref{fig:sensitivity} shows the upper and lower limits calculated via the methods described in this and the previous section. The blue triangles show the upper limits of flux density values of transient sources we would be sensitive to at various time-scales, based on the filters described in Table \ref{tab:filtering_radii}. The red triangles show the lower limits on the transient search based on the simulations (the 2 minute snapshot value is inferred from the 8 second and 1 hour subtracted images). The green shaded region between the blue and red markers indicates the flux density values the method presented in this paper is sensitive to. Furthermore, Figure \ref{fig:sensitivity} shows black stars that, per time-scale, indicate the rms noise level per image (see Section \ref{sec:results_image_quality}) times the detection threshold (defined in Section \ref{sec:methods_detection_thresholds}). This gives the theoretical lower limit of sources we could detect in the subtracted images per time-scale. 
Finally, Figure \ref{fig:sensitivity} shows the $1/\sqrt{t}$ relation that is used to derive the lower limit to the search in the 2 minute shapshots, based on the 8 second and 1 hour subtraction images.

Finally, we want to point out that in some cases the subtraction images could slightly underestimate the flux of a transient candidate. For example, a 40 mJy transient that is 'on' for one hour throughout the 8 hour observation will show up as a ($40/8=$) 5 mJy source in the deep image. This means that a 5 mJy source will be subtracted out at the location of the transient during the source subtraction. This will lead to a transient flux density that is slightly lower than the true flux density, and additionally, it will create negative subtraction artefacts at times when the transient is off. We do not consider this effect in detail in our search, as the main goal of the subtracted images is to identify transient sources, characterization will follow with additional imaging.

\section{Results} \label{sec:results}
We apply the methods outlined above to 58 pointings of LoTSS. These pointings correspond to the sky area covered by LoTSS-DR1 \citep{shimwell2019lofar}, but we note that the data is reprocessed using the LoTSS-DR2 approach \citep{shimwell2022lofar}.

\subsection{Subtracted images quality} \label{sec:results_image_quality}
\begin{figure}
    \centering
    \includegraphics[width=\linewidth]{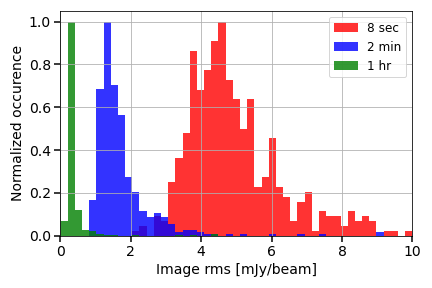}
    \caption{Rms noise in a sample of 500 subtracted images per time-scale. The rms distributions for the 8 second snapshot images, 2 minute timeslices and 1 hour images are shown from left to right respectively. Only the inner the innermost circle with radius $1.5 ^{\circ}$ is used to construct these distributions.}
    \label{fig:rms_per_time-scale}
\end{figure}

Figure \ref{fig:rms_per_time-scale} shows the rms noise for the subtracted images for a subset of 500 images per time-scale.  The rms distributions for the 8 second images, 2 minute timeslices and 1 hour time-scale are shown from left to right respectively. No sigma clipping is performed on these distributions, so they also contain source pixels. The rms noise is around $~4.6$ mJy/beam for the 8 second snapshots, $~1.3$ mJy/beam for the 2 minute snapshots and $~0.3$ mJy/beam for the 1 hour snapshots. 
Multiplying these numbers with the detection thresholds applied at the various time-scales (see Table \ref{tab:detection_thresholds}) gives a lower limit on the flux density of the faintest transient that we could detect at each time-scale. In Figure \ref{fig:sensitivity} these numbers are indicated by the black stars and they are in good agreement with the detection thresholds determined with the simulations in Section \ref{sec:simulations}. These lower limits seem to probe slightly deeper than the simulations, but that is mainly an effect of the quite crude steps in flux density in our simulations, see Figure \ref{fig:sim_results}. Furthermore, these numbers are in good agreement with the expected scaling of the sensitivity with time as $1/\sqrt{t}$.

\subsection{Method efficiency}
\begin{figure}
    \centering
    \includegraphics[width=\linewidth]{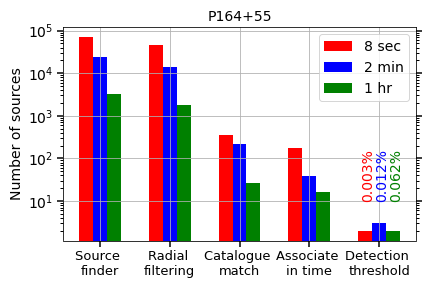}
    \caption{Number of sources left after each step in the analysis for the P164+55 field for different snapshot time-scales. The numbers above the final step show the percentage of sources left for visual inspection compared to the number of sources found by the source finder. }
    \label{fig:nof_sources_per_step}
\end{figure}

Figure \ref{fig:nof_sources_per_step} shows the number of sources left after each step in the analysis for the P164+55 field for different snapshot time-scales. The numbers above the final step show the percentage of sources left for visual inspection compared to the number of sources found by the source finder in the first step. The steps described in the methods section are able to filter out $\sim 0.003 - 0.06 \%$ of sources as potential transient candidates. This corresponds to in total $\mathcal{O}(10)$ candidates per field, which is feasible but tedious for visual inspection.

%


\subsection{Transient candidates}
After processing 58 fields that correspond to the fields presented in the first data release of LoTSS \citep{shimwell2019lofar},  8 transient candidates are identified. The subtracted images where the sources are detected are listed in Appendix \ref{app:all_candidates}.

In the following sections, we decide to consider the transient candidates shown in Figure \ref{fig:appendix_candidate_P_8sec}, \ref{fig:appendix_candidate_P_128sec} and \ref{fig:appendix_candidate_P35Hetdex10}. This is because these are the brightest candidates and the candidates in the subtracted images have a shape similar to the dirty beam, similar to the simulations (See Figure \ref{fig:8sec_sim}). As the candidates in Figures \ref{fig:appendix_candidate_P_8sec} and \ref{fig:appendix_candidate_P_128sec} are two instances of the same candidate detected in different time-scale, these two detections are discussed jointly in more detail in Section \ref{sec:results_candidate1}. The candidate in Figure \ref{fig:appendix_candidate_P35Hetdex10} is discussed in more detail in Section \ref{sec:results_candidate2}.


 \section{Discussion} \label{sec:discussion}
In this Section we discuss the implications of our results. First, in Sections \ref{sec:results_candidate1} and \ref{sec:results_candidate2} we discuss in more detail the origin of the two transient candidates we identified. To study the candidates in more detail the sources were imaged with primary beam correction and cleaning using the direction-independent data products. In the next sections, we discuss the upper limits we can place on the transient surface density. Finally, we discuss how this work could be extended to look for variable sources.


\subsection{Transient candidate 1} \label{sec:results_candidate1} 
The first transient candidate found in this study is presented in Figures \ref{fig:appendix_candidate_P_8sec} and \ref{fig:appendix_candidate_P_128sec}. There is a bright dirty-beam shaped source in the subtracted image that cannot be associated with any source in the deep image. The source is detected in three consecutive 8 second subtracted images with signal-to-noise ratio increasing from 6.1 to 7.7 and 8.8. Additionally, the source is identified in the 2 minute snapshot that encompasses this 24 second interval with a signal-to-noise ratio of 7.3.

\subsubsection{Validity checks}
To make sure that this transient source is real and not an imaging artefact, we perform some additional checks. In contrast to the artefacts discussed in Appendix \ref{app:psf_artefacts} and the transient candidate discussed in Section  \ref{sec:results_candidate2}, the location of the source does not change in the subtracted images, as the brightest pixel is identified to be the exact same pixel in each of the three images. This is what we expect to be the case for an astrophysical transient. The distance from the transient location to the center of the beam $0.82 ^{\circ}$, most artefacts as discussed in Appendix \ref{app:psf_artefacts} were found close to the edge of the search radius ($1.5 ^{\circ}$). Furthermore, recreating the subtracted images with additional padding and slightly different imaging settings does not make the transient candidate disappear. This assures us that our transient candidate is not one of the PSF artefacts discussed in Appendix \ref{app:psf_artefacts}.

Additionally, Figure \ref{fig:C1_noise_properties} shows the noise distribution of the particular subtracted images where the transient is identified compared to the rms noise in all 8 second subtracted images of this pointing. Figure \ref{fig:C1_noise_properties} shows that these subtracted images do not show increased noise compared to the full set of images. Comparing Figure \ref{fig:C1_noise_properties} to Figure \ref{fig:rms_per_time-scale} reveals that the noise in this particular field is on the high side compared to 8 second subtracted images of other fields. This is due to poorer ionospheric conditions during this observations, which we will shortly discuss in the next section.

\begin{figure}
    \centering
    \includegraphics[width=\linewidth]{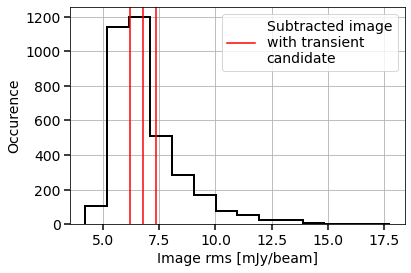}
    \caption{RMS noise in the 8 second subtracted images of field where the transient candidate was found. The coloured vertical lines show the rms noise in the particular subtracted images where the transient candidates were identified.}
    \label{fig:C1_noise_properties}
\end{figure}

\subsubsection{Reimaging}
Now that a viable transient candidate is identified, additional imaging is performed to fully characterize the source. Using the publicly available products presented in \cite{shimwell2022lofar}, we reimage the direction-independent calibrated visibilities with particular interest around the time when the transient candidate is found. In this process, the size of a pixel is decreased in order to get a more accurate position measurement, some of the shortest baselines are cut out to get rid of extended emission and deep cleaning is performed. The most important imaging parameters are presented in Table \ref{tab:wsclean_DI_P}. Now that cleaning and a primary beam correction are applied, a more accurate estimate of the flux of the transient can be made.

\begin{table}
\begin{tabular}{ll} \hline
Setting                    & Value        \\ \hline
Number of clean iterations & 150000            \\
Auto-mask                   & 2.5 \\
Auto-threshold              & 0.5 \\
minuv-l                     & 80 \\
channels-out                    & 3 \\
Size of image (pixels)     & 4400         \\
Size of one pixel (arcsec) & 3            \\
Briggs weighting           & -0.5      \\ 
Interval                   & 2612-2621 \\
Intervals out              & 9 \\
Padding                    & 1.4 \\ 
apply primary beam          & True \\ \hline
\end{tabular}
\caption{\label{tab:wsclean_DI_P} The {\sc WSClean} settings used to reimage the transient candidate presented in Section \ref{sec:results_candidate1}.}
\end{table}

\begin{figure}
    \centering
    \includegraphics[width=0.9\linewidth]{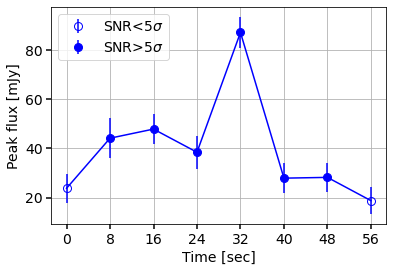}
    \caption{Peak flux density of transient candidate 1 as a function of time. Different markers indicate different levels of signal-to-noise ratio of the detection.}
    \label{fig:C1_lightcurve}
\end{figure}

Figure \ref{fig:C1_lightcurve} shows the peak flux density as a function of time based on the reimaged 8-second snapshots. The spectral index at the peak of the flare is negative $\alpha\approx -1.0$ (for $S\propto \nu^{\alpha}$).

Separate images were made for the Stokes I and V components of the signal, and no circularly polarized flux was detected. The data quality is insufficient to perform the frequency slicing necessary for QU fitting. 
It would be better to perform an in-depth study of the polarization properties using the direction-dependent self-cal solutions. Unfortunately, this pointing has particularly bad ionospheric conditions and we were unable to improve the calibration compared to the LoTSS-DR2 images. Improved calibration would not only be useful in studying the polarization properties but also in understanding the spectrum of the flare, as this procedure would increase the SNR. To this end additional observations under more favourable ionospheric conditions are necessary.

\subsubsection{Future work}
We leave it to a future paper \citep{deruiter2024inprep} to determine the true nature of this transient source. To this end, we will analyse additional observations of this field. The 8 second integration time of the data and relatively low signal-to-noise ratio (preventing us from making many frequency slices), hampers a dispersion analysis for the flare. If an intrinsically short duration signal is dispersed to $6\cdot 8 = 48$ seconds (taking only the detections with SNR$>5$), this points to a DM of roughly
\begin{equation}
    \rm{DM} = \frac{1.2 \cdot 10^5 \cdot \tau_{\rm{DM}} \cdot \nu^3}{B} = \frac{1.2 \cdot 10^5 \cdot 48  \cdot 0.144^3}{48} = 358 \; \rm{pc}\, \rm{cm}^{-3}
\end{equation}
with $\tau_{\rm{DM}}$ the dispersion measure smearing in seconds, $\nu$ the central frequency in GHz, and B the bandwidth in MHz. A DM of $358 \; \rm{pc}\, \rm{cm}^{-3}$ would make this an extragalactic source \citep{cook2023frb}. \\

Another possibility is that the radio pulse produced by this source intrinsically has a duration of half a minute to a minute. Recently galactic sources of this type have been discovered by \cite{hurley2022radio} and \cite{hurley2023long}.

\subsection{Transient candidate 2} \label{sec:results_candidate2} 
The same reimaging process that was outlined in Section \ref{sec:results_candidate1} is repeated for the transient candidate presented in Figure \ref{fig:appendix_candidate_P35Hetdex10}.



\begin{figure*}
    \centering
	\includegraphics[width=\textwidth]{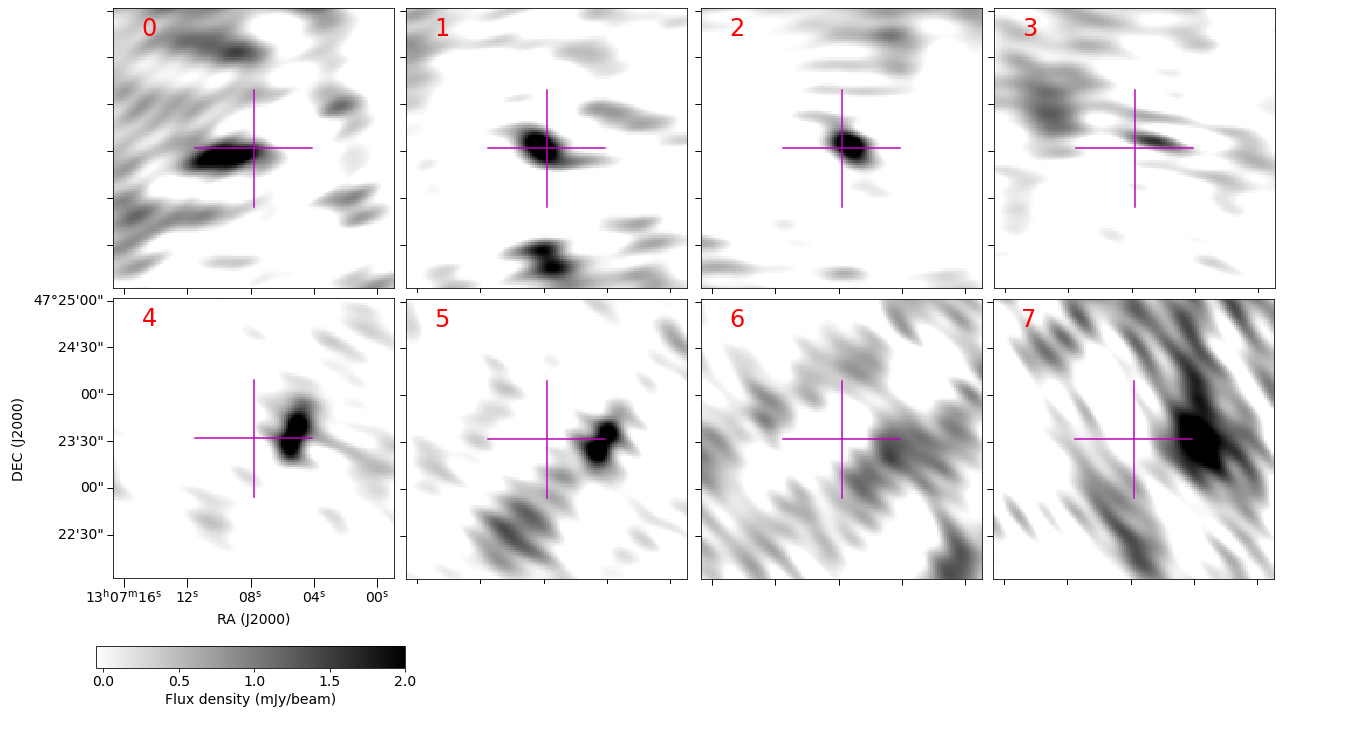}
    \caption{Insets at the location of transient candidate 2 in the snapshot images, which are created  with primary beam correction and cleaning using the direction dependently calibrated data. The crosshair is fixed at the same location in all panels. The panels show the 1 hour snapshots throughout the 8 hour observation. The transient candidate moves throughout the observation.}
    \label{fig:P35Hetdex10_candidate_insets}
\end{figure*}

\begin{figure}
    \centering
    \includegraphics[width=\linewidth]{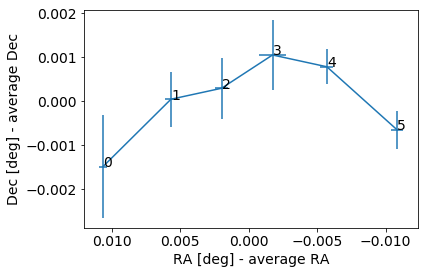}
    \caption{The right ascension and declination of the transient candidate 2 in degrees minus its average location. The numbers indicate the snapshot number. This plot only contains the snapshots where the source was detected at $>5\sigma$ above the rms noise. }
    \label{fig:P35Hetdex10_candidate_trajectory}
\end{figure}

The new images reveal that the location of the transient candidate changes throughout the observation. Figure \ref{fig:P35Hetdex10_candidate_insets} shows insets of the transient candidate in the new images. The crosshair is fixed at the same location in all panels. The panels show the 1 hour snapshots throughout the 8 hour observation. From these images, it is clear that the transient candidate moves throughout the observation. This effect is quantified in Figure \ref{fig:P35Hetdex10_candidate_trajectory}, which shows the right ascension and declination of the source minus its average location. The numbers indicate the snapshot number. This plot only contains the snapshots where the source was detected with $>5\sigma$. The average position of the source is $(196.778^\circ, 47.392^\circ)$. Studying the deep LoTSS image (right panel in Figure \ref{fig:appendix_candidate_P35Hetdex10}) carefully, there is an arc-shaped artefact visible in the 8 hour average image. This arc roughly corresponds to the path of the source, see Figure \ref{fig:P35Hetdex10_candidate_trajectory}.

Including the marginal detection of the last hour snapshot, this implies that our candidate sources moves about 1 arcmin on the sky over 8 hours. This movement is clearly different from the slight shift in position that other sources experience from snapshot to snapshot. Especially in right ascension the movement of this source is $8\sigma$ away from the jitter that other sources in the field experience. Furthermore, most sources seem to move in a random direction from image to image, but our transient candidate seems to follow a trajectory, it does not move back toward its original position. If we assume this is a real astrophysical source, this movement implies a high proper motion and/or an extremely nearby object. The displacement places the source within roughly one parsec, where it would travel with the speed of light. This suggests that the source, if astrophysical, is most likely within our Solar system. At an average position of $(196.778^\circ, 47.392^\circ)$ this source lies approximately $70^{\circ}$ above the Galactic Plane and about $50^{\circ}$ above the ecliptic. 

Additional checks and tests that were performed do not provide any further insight. The emission is broadband, but hints to be brighter towards lower frequency. This seems to exclude most reflected signals and satellites. There is no significant stokes Q,U or V detection. Due to the position shift it is difficult to match the source to other catalogues. No minor planets were found at the location of our transient using \url{https://minorplanetcenter.net/cgi-bin/checkneo.cgi} around the observation time of this field (2014-07-14 14:00:00 UTC). A highly speculative origin for this type of emission is an asteroid or comet reflecting emission from the Sun. Radio reflection tomography has been proposed as a technique to study the interior of asteroids and comets (see eg. \cite{safaeinili2002probing}) and the Sun has been used as a radio source to use a similar reflection technique to probe the thickness of glaciers on Earth \citep{peters2021glaciological}. During the observation of this field, there was no extreme solar activity\footnote{\url{https://solen.info/solar/old_reports/2014/july/20140715.html}}.

A final note we would like to make is that the source trajectory (Fig. \ref{fig:P35Hetdex10_candidate_trajectory}) is quite similar to the trajectories that we observe as a result of artefacts close to the edge of the image, such as for example shown in Figure \ref{fig:example_trajectory_psf_artefact}. However, all the 'fake' transient sources that were found to show such a trajectory were detected quite close to the edge of the image, and the sources disappeared altogether when reimaging with additional padding. Following this procedure did not make the transient source discussed here disappear. 

The most likely explanation is that this source is an imaging artefact. The arc-shaped artefact present in the LoTSS deep image is similar to artefacts introduced by facet calibration due to aliasing or flagging. We note that neither increased padding nor turning off flagging removed this source from the images. An extensive investigation into the origin of this artefact is beyond the scope of this work.

In conclusion, we find a source in the data that passes all tests we do in our analysis, and proper reimaging shows that there is indeed a transient point source in the data. However, we are currently unaware of an astrophysical process that would explain a source that shows broad-band transient radio emission and shows a high proper motion, which implies it is located in the Solar system. The most likely explanation for this source is that is an imaging artefact. If this is an astrophysical source we expect to find more similar sources in our follow-up study ($\sim13$), where we plan to repeat this study for a larger sky area.

\subsection{Transient rates} \label{sec:trans_limits}
Following \cite{rowlinson2016limits} we calculate the transient surface density limit using Poisson statistics via 
\begin{equation}
P(X=n) = \frac{\left(\rho \cdot \Omega_{total}\right)^n e^{-\rho \cdot \Omega_{total}}}{n!},
\label{eq:trans_rate}
\end{equation}
where $\Omega_{total}$ is the total area surveyed at a certain time-scale, $\rho$ is the surface density limit and $P$ is the confidence interval. In case no transient candidates are detected this equation reduces to $P(X=0) = e^{-\rho \Omega_{total}}$ which allows one to define an upper limit on the transient surface density. Following \cite{bell2014survey}, we use $P=0.05$ to give a 95 percent confidence limit. The total sky area surveyed in this work is summarized in Table \ref{tab:transient_surface_density}. $\Omega_{total}$ shows the naive calculation of the surveyed sky area, ie. the number of fields, 58, times the number of images per time-scale, N. However, in this work we perform cuts around bright sources that reduce the effective sky area we search. $\Omega_{corr}$ shows the total sky area surveyed while taking into account the area we cut out. As we do not find any transients in the 1 hour snapshots, an upper limit is calculated for this time-scale. One transient candidate is found in 2 minute and 8 second snapshots, and the corresponding transient surface densities are shown in the final column of Table \ref{tab:transient_surface_density}.


\begin{table}
\begin{tabular}{|l|l|l|l|l|}
\hline
Time scale  & N  & $\Omega_{total} [\text{deg}^{2}]$ & $\Omega_{corr} [\text{deg}^{2}]$ & $\rho \; [\text{deg}^{-2}]$ \\ \hline
1 hour      & 8     & 3280      & 3210    & $<4.0 \cdot 10^{-4}$         \\
2 min     & 225   & 92245     & 92218   & $5.7 \cdot 10^{-7}$         \\
8 sec       & 3600  & 1475920   & 1475908 & $3.6 \cdot 10^{-8}$        \\ \hline
\end{tabular}
\caption{The transient surface density for various time-scales. \label{tab:transient_surface_density}}
\end{table}


\begin{figure*}
    \centering
	\includegraphics[width=\textwidth]{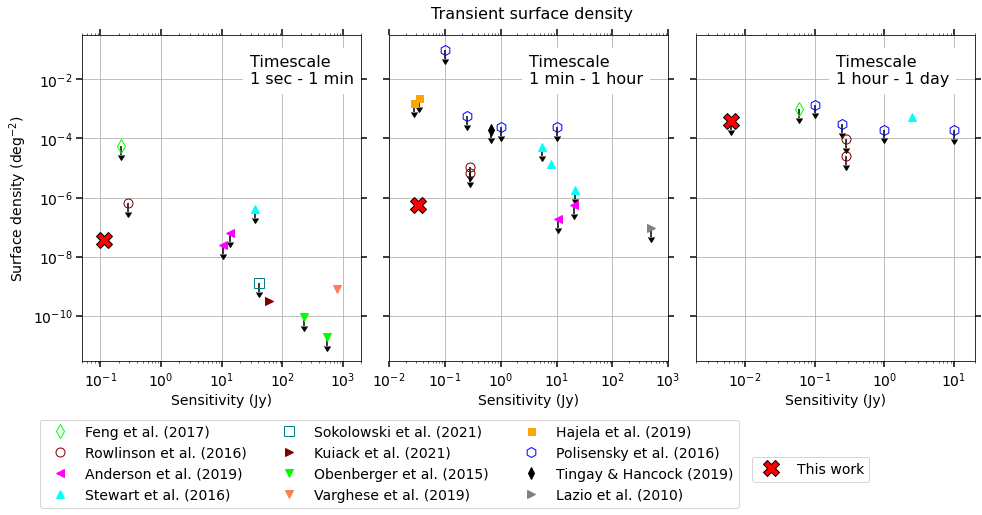}
    \caption{Limits on the transient surface density from this study compared to previously published results below 340 MHz. The three panels show studies probing different transient time-scales. The result presented in this paper is shown as a red cross. Other results are from \protect \cite{lazio2010surveying, obenberger2015monitoring, stewart2016lofar, anderson2019new, varghese2019detection, kuiack2021aartfaac, hajela2019gmrt, tingay2019multi, polisensky2016exploring, rowlinson2016limits,feng2017matched, sokolowski2021southern}. 
    \label{fig:transient_rates_new}}
\end{figure*}

Figure \ref{fig:transient_rates_new} shows our new result (red cross) compared to other results in the literature. The figure consists of three panels probing three different transient duration times. We compare our results to the most constraining studies \textit{below 340 MHz} and find that our results are probing the lowest sensitivities to date at all transient time-scales. The structure of this plot was taken from \cite{murphy2017search}, but a more up-to-date sample of the most constraining studies was compiled using an overview\footnote{\url{http://www.tauceti.caltech.edu/kunal/radio-transient-surveys/index.html}} from \cite{mooley2016caltech}. Markers with a downward pointing arrow represent upper limits. We choose to show the results of our study as datapoints at a fixed sensitivity, instead of as a curve showing the transient surface density as a function of sensitivity by including a larger portion of images with higher rms noise values (see for example \cite{rowlinson2022search}). We opt not to do this as we show that the sensitivity is quite uniform across the image (see Figure \ref{fig:sim_results}). From Figure \ref{fig:transient_rates_new} it is clear that the transient source presented in Section \ref{sec:results_candidate1} could not have been detected by previous studies, as those did not probe sufficient sky area and/or lacked sensitivity compared to our search. 

\subsection{Transient rates at higher frequencies}
Many radio telescopes have typical observing frequencies of around 1.4 GHz. Therefore, in this section, we compare the transient surface density values from this work (at 144 MHz) to some of the most constraining studies at 1.28-1.4 GHz. From the papers below, we distill the information necessary to calculate the transient surface density following Equation \ref{eq:trans_rate}.

\cite{chastain2023commensal} use MeerKAT to perform a commensal search for transients within the ThunderKAT program \citep{fender2017thunderkat}. Images are created with integration times of 4 hours, 15 minutes and 8 seconds, with median noise levels of 10, 30 and 176 $\mu$Jy respectively (Table 2 in \cite{chastain2023commensal}). Furthermore, detection thresholds of 5.3, 5,7, and 6.4$\sigma$ are used. These observations are taken at 1.28 GHz and each image has a size of at least 2.8 by 2.8 $^{\circ}$. In total 28 4-hour images, 406 15-minute images, and 43964 8-second images were created. The transient surface density resulting from this study is calculated using Eqn. \ref{eq:trans_rate} and shown in Table \ref{tab:trans_surf_dens_high_freq}. We also note the work by \cite{anderson2022rapid}, who find a flare stare in a commensal search for transients in MeerKAT observations, but this search does not go as deep (in the number of epochs) as the work by \cite{chastain2023commensal}.

Additionally, we compare our results to the work by \cite{fijma2023new}, as their imaging strategy is similar to ours. They perform a transient search in MeerKAT data around the NGC 5068 field, creating snapshot images using the subtraction imaging methods as explained in Section \ref{sec:methods_sky_model_subtraction}. Furthermore, the time-scales used for imaging are identical to what was used in this study. The transient surface density limits from \cite{fijma2023new} are summarized in Table \ref{tab:trans_surf_dens_high_freq}.

Finally, \cite{wang2023radio} perform a transient search on 15-minute time-scales using the Variable and Slow Transient survey using the Australian SKA Pathfinder (ASKAP; \cite{hotan2021australian}). They find 3 single flare transients in 754 images at 1.4 GHz. Again, this study is of particular interest as a similar source subtraction method is used to create images. The size of an individual image is 2.1 x 2.1$^{\circ}$ and the median rms noise in an individual image is around 0.2 mJy $\rm{beam}^{-1}$. A detection threshold of $6\sigma$ was used. The transient surface density resulting from this study is calculated using Eqn. \ref{eq:trans_rate} and shown in Table \ref{tab:trans_surf_dens_high_freq}. We also note the work by \cite{dobie2023radio} but instead, choose to include just the work by \cite{wang2023radio}, as the latter provides a more constraining result.

\begin{table}
    \centering
    \begin{tabular}{l|l|l|l|l|l}
    S & $S_{\rm{144 \; MHz}}$    & Transient surface        &    Time-scale       &  Frequency      &  Ref. \\
     (mJy)      &  (mJy)   & density ($\rm{deg}^{-2}$)  &  & (GHz) &  \\ \hline
    $0.053$     &  $0.29$  & $< 1.4\cdot 10^{-2}$ & 4 hours & 1.28 &  [1] \\ 
    $3.9 $      & $23.7$   & $< 3.2 \cdot 10^{-2}$  & 1 hour        & 1.4 & [2] \\
    -      & $6.3$   & $<4.0 \cdot 10^{-4}$  & 1 hour        & 0.144 &  \\ 
    -      & $6.3$   & $<1.6 \cdot 10^{-3}$  & 1 hour        & 0.144 &  [**]\\  \hline
    $0.17$      &  $0.94$  & $< 9.4\cdot 10^{-4}$ & 15 min & 1.28 & [1] \\
    $1.2$       & $7.28$   & $ 3.2 \cdot 10^{-4}$   & 15 min          & 1.4 & [3]\\ 
    $19.2$      & $117$    & $< 1.1 \cdot 10^{-3}$  & 2 min   & 1.4 & [2] \\ 
    -     & $30$    & $5.7 \cdot 10^{-7}$  & 2 min   & 0.144 &  \\ 
    -     & $30$    & $4.3 \cdot 10^{-6}$  & 15 min   & 0.144 &  [**]\\ \hline
    $1.13$      &  $6.25$  & $< 8.7\cdot 10^{-7}$ & 8 sec & 1.28 & [1] \\
    $56.4 $     &  $342$   & $< 6.7 \cdot 10^{-5}$  & 8 sec     & 1.4 & [2] \\
    -     &  $113$   & $3.6 \cdot 10^{-8}$   & 8 sec     & 0.144 &  \\
    \end{tabular}
    \caption{Comparison of the transient surface density as found in this study, compared to transient searches at higher frequency. The sensitivity at the frequency of the original study is extrapolated to 144 MHz assuming a spectral index $\alpha=-0.783$, where $S_{\nu}\propto \nu^{\alpha}$. References [1]:\protect \cite{chastain2023commensal}, [2]:\protect \cite{fijma2023new}, [3]:\protect \cite{wang2023radio}, [**]: results from this study that have been extrapolated to longer time-scales to allow a direct comparison with other studies.}
    \label{tab:trans_surf_dens_high_freq}
\end{table}

Table \ref{tab:trans_surf_dens_high_freq} shows an overview of the transient surface density values as presented in the studies detailed above. The first column shows the detection sensitivity as originally mentioned in the study, whereas the second column shows how that sensitivity would translate to 144 MHz assuming a spectral index $\alpha=-0.783$, where $S_{\nu}\propto \nu^{\alpha}$. This is the average spectral index of LoTSS sources when compared to NVSS at 1.4 GHz (see Figure 18 in \cite{shimwell2022lofar}). The other columns show the transient surface density, time-scale, observing frequency and reference to the study respectively. Each block of Table \ref{tab:trans_surf_dens_high_freq} compares studies that have been performed at a similar time-scale. The final rows in each block show the results from our work (as calculated in Table \ref{tab:transient_surface_density}). For the hour and minute time-scale we additionally extrapolate our results to the exact time-scales probed in \cite{chastain2023commensal} and \cite{wang2023radio} to make a fair comparison. This is done by multiplying the transient surface density with a factor of 4 and 7.5 for hours and minutes time-scales respectively. 

From Table \ref{tab:trans_surf_dens_high_freq} it is evident that even when assuming a flat spectral index the detection sensitivity in our work is of the same order of magnitude as in \cite{fijma2023new}. Extrapolating the detection sensitivity in \cite{fijma2023new} to 144 MHz shows that our study probes roughly an order of magnitude deeper at hours and minutes time-scales, and roughly three times deeper at second time-scales. Additionally, our work provides much deeper limits on the transient surface density at all time-scales. 

The last row in the minutes time-scale block of Table \ref{tab:trans_surf_dens_high_freq} shows our 2 minute time-scale results extrapolated to 15 minutes. Our results are of similar sensitivity compared to \cite{wang2023radio}, but the transient surface density is much more constraining.
Finally, our study is less sensitive than the extrapolated sensitivity presented in \cite{chastain2023commensal} for the hours, minutes and seconds time-scales, but in contrast to \cite{chastain2023commensal} we do find a transient source in the 2 minute and 8 second snapshots, most likely due to our much larger observing time. Therefore, our transient surface density rates are more constraining. 

A curiosity we would like to point out regarding Table \ref{tab:trans_surf_dens_high_freq} is that \cite{wang2023radio} do find 3 interesting flaring transient sources at a sensitivity in between ours and \cite{chastain2023commensal}, which is surprising. The discrepancy between \cite{wang2023radio} and \cite{chastain2023commensal} could be due to the larger sky area probed by \cite{wang2023radio}, but there is also almost an order of magnitude sensitivity difference, which is expected to play a role. Additionally, the fact that \cite{wang2023radio} find three sources at minutes time-scales, compared to our one source implies that we might probe different transient source populations at 144 MHz and 1.4 GHz. This could for example be due to a synhrotron self-absorbption break in the spectrum that causes some of the transients found at 1.4 GHz to be fainter at 144 MHz, causing us to miss them in our transient searches. Finally, it would be good to in the future thoroughly compare methods with \cite{wang2023radio} as there might be some intrinsic differences to our search, leading to different results.

\subsection{Stewart et al. transient}

To date, only a handful of transient surveys at low radio frequencies (<1 GHz) have detected transient sources. Examples include \cite{hyman2009gcrt, bannister201122, jaeger2012discovery} at time-scales of days to months, which have low detection probability in our survey because the largest snapshot time we use in 1 hour. More relevant is the bright (15-25 Jy) transient identified at 60 MHz by \cite{stewart2016lofar}, with a duration of a few minutes. In the next section, we calculate the expected number of observed Stewart-like transients in our survey, as a function of spectral index. Using a conservative flux of 15 Jy and survey frequency of 60 MHz for the \cite{stewart2016lofar} survey, we can calculate the flux we expect to observe at our survey frequency (144 MHz), assuming some spectral index $\alpha$, where $S\propto \nu^\alpha$. 

The sensitivity of this survey can be scaled to \cite{stewart2016lofar} survey using $N\propto S^{-3/2}\cdot \Omega$ as expected for an isotropic homogeneous distribution of sources throughout a flat space, where $S$ is the sensitivity of the survey and $\Omega$ is the total surveyed area. The expected number of Stewart-like transients in our survey is then equal to the ratio of the surveyed area in our survey and the Stewart survey times the detection sensitivity at minutes time-scale in our survey divided by the expected flux of the Stewart-like transient at the observing frequency of our survey, to the power $-3/2$:
\begin{equation}
    N = \frac{\Omega_{\rm{LoTSS}}}{\Omega_{\rm{Stewart}}}\left(\frac{S_{\rm{survey, minutes}} \; \rm{Jy}}{15 \; \rm{Jy} \cdot \left(\frac{144 \; \rm{MHz}}{60 \; \rm{MHz}}\right)^{\alpha}}\right)^{-3/2}
\end{equation}

For the detection sensitivity of our survey at minutes time-scale we use a value of $\sim 30$ mJy, by extrapolating our simulation (see Section \ref{sec:simulations}). Note that we use a flux density of 15 Jy for the Stewart et al. transient, in contrast to \cite{rowlinson2016limits} where the detection limit of the survey is used. Table \ref{tab:stewart_compare} gives the expected number of transients on the minutes time-scale, as a function of spectral index, scaled from the \cite{stewart2016lofar} transient detection.

\begin{table}
\begin{tabular}{l|ll} \hline
Spectral index & Number predicted  & Null detection probability     \\  \hline
$-4$ & 16 & $8.8\cdot 10^{-8}$ \\
$-5$ & 4.4 & $1.3\cdot 10^{-2}$ \\
$-6$ & 1.17 & 0.31 \\
$-7$ & 0.32 & 0.73  \\
$-8$ & 0.085 & 0.92 \\ \hline
\end{tabular}
\caption{\label{tab:stewart_compare} The expected number of transients on the minutes time-scale, as a function of spectral index, scaled from the \protect\cite{stewart2016lofar} transient detection.}
\end{table}

Table \ref{tab:stewart_compare} shows that we can rule out spectral indices $\geq -5$ at 95 percent confidence for the \cite{stewart2016lofar} transient, based on our extrapolated survey sensitivity estimate. Combining this result with \cite{rowlinson2016limits}, we conclude that either the transient rate derived in \cite{stewart2016lofar} is too high or that this transient event had an extreme spectral index.

\subsection{Variability}
In this work we do not study variable sources, but the subtracted images presented in this work are suitable to study variable sources. A variable source would show up in the subtracted images just like the example in Figure \ref{fig:sub_image_art_example}. Either a positive artefact would be present if the source is brighter than the sky model flux at the subtracted image snapshot time, or a negative artefact if the source is dimmer than the sky model flux at the specific snapshot time. In our analysis, we disregard these sources as they are associated with a LoTSS catalogue source. However, one could imagine setting a threshold where very bright variable source artefacts would be kept for further analysis. Highly variable sources can be used to study interstellar scintillation, see for example \cite{anderson2022rapid}. Therefore, an interesting future extension to the work presented here would be to include (highly) variable sources in our analysis.
Additionally, our work excludes transient sources associated with bright radio sources or galaxies. A transient that lies in close proximity to radio sources with a flux density that falls within the filter limits described in Table \ref{tab:filtering_radii} will be excluded for our search. This biases our search against transients occurring in galaxies with high radio flux.

\section{Conclusion \& Outlook} \label{sec:conclusion}
In this paper, we have presented the results of a search for transient sources at 144 MHz using the LOFAR Two-metre Sky Survey \citep{shimwell2022lofar}. The search covers mostly extragalactic sky areas and covers $410 \; \rm{deg}^2$ of sky. This search was performed on snapshot images with integration time-scales of 8 seconds, 2 minutes and 1 hour, splitting the 8 hour pointings in roughly 3600, 225 and 8 snapshots. In order to create these snapshot images we use a new approach where we search for transients in the subtracted images. These are created by subtracting the full (8-hour) sky model from the visibilities of the snapshot and imaging those visibilities without cleaning or primary beam correction. This process greatly reduces the otherwise computationally very expensive imaging step. Afterwards, source finding, filtering and source catalogue matching steps are applied to find the transient sources. One transient candidate is identified, but follow-up is necessary to determine its true nature. This work identifies the lowest transient surface density at time-scales of seconds to hours at the highest sensitivity to date. In the future, this method will be applied to the second data release of LoTSS. This will increase the number of processed pointings from 58 in this study to 841. 

Another approach that might be explored in the future is the use of a different source finding technique, more suitable for subtracted images. The source finder used in this work already accounts for complicated noise structures throughout the images, but subtracted images introduce other difficulties. Since the subtracted images are not cleaned, each source appears as a blob consisting of multiple components. Using the \textsc{LPF} sourcefinder all these components are identified as individual sources. A source finder that would automatically identify these components to be part of the same uncleaned source would simplify the filtering steps afterwards and speed up the process as there would be fewer sources to consider. Secondly, if the techniques presented in this paper would be applied to future studies where many transients are identified, an automated cross-matching to other catalogues might be useful in order to determine the origin of the transient emission. Finally, future datasets or surveys will potentially need a more stringent catalogue match or filtering scheme to reduce the number of transient candidates left for visual inspection, but we find the current filter scheme gives a good trade-off between not disregarding transient candidates too soon and time needed for visual inspection.

\section*{Acknowledgements}

This research made use of \textsc{astropy} \citep{astropy:2013, astropy:2018} for FITS file handling and coordinate matching and \textsc{matplotlib} \citep{hunter2007matplotlib} was used to create plots.

LOFAR is the Low Frequency Array designed and constructed by ASTRON. It has observing, data processing, and data storage facilities in several countries, which are owned by various parties (each with their own funding sources), and which are collectively operated by the ILT foundation under a joint scientific policy. The ILT resources have benefited from the following recent major funding sources: CNRS-INSU, Observatoire de Paris and Université d'Orléans, France; BMBF, MIWF-NRW, MPG, Germany; Science Foundation Ireland (SFI), Department of Business, Enterprise and Innovation (DBEI), Ireland; NWO, The Netherlands; The Science and Technology Facilities Council, UK; Ministry of Science and Higher Education, Poland; The Istituto Nazionale di Astrofisica (INAF), Italy.

This research made use of the Dutch national e-infrastructure with support of the SURF Cooperative (e-infra 180169) and the LOFAR e-infra group. The Jülich LOFAR Long Term Archive and the German LOFAR network are both coordinated and operated by the Jülich Supercomputing Centre (JSC), and computing resources on the supercomputer JUWELS at JSC were provided by the Gauss Centre for Supercomputing e.V. (grant CHTB00) through the John von Neumann Institute for Computing (NIC).

This research made use of the University of Hertfordshire high-performance computing facility and the LOFAR-UK computing facility located at the University of Hertfordshire and supported by STFC [ST/P000096/1], and of the Italian LOFAR IT computing infrastructure supported and operated by INAF, and by the Physics Department of Turin university (under an agreement with Consorzio Interuniversitario per la Fisica Spaziale) at the C3S Supercomputing Centre, Italy.

This research was funded through project CORTEX (NWA.1160.18.316), in research programme NWA-ORC, financed by the Dutch Research Council (NWO).

\section*{Data Availability}
The data and data products presented in this paper are available in a reproduction package via Zenodo, at \url{https://dx.doi.org/10.5281/zenodo.10118363}.



\bibliographystyle{mnras}
\bibliography{ms} 




\appendix

\section{Mathematical description source subtraction}\label{app:maths}
Below we describe the sky model subtraction process in a more mathematical manner. An interferometer measures an approximate Fourier transform of the sky intensity, instead of directly measuring the sky intensity $I(l,m)$. The intensity and visibility are defined as a function of viewing direction cosines $(l,m)$. For a baseline consisting of antennas $i$ and $j$, the perfect response to all visible sky emission for a single time instance and frequency is given by the idealized Radio Interferferometric Measurement Equation (RIME), see eg. \cite{smirnov2011revisiting}. The RIME below does not include the Jones matrices that describe the direction-dependent and direction-independent calibration effects. The visibilities for a baseline consisting of antennas $i$ and $j$ are defined as
\begin{equation}
V_{ij} = \int \int I(l,m) e^{-2\pi i \left[u_{ij}l + v_{ij}m + w_{ij}(n-1) \right]} \frac{dl dm}{n}
\label{eqn:RIME}
\end{equation}
where $i=\sqrt{-1}$, $n=\sqrt{1-m^2-l^2}$, $u_{ij}$ and $v_{ij}$ are baseline coordinates in the UV plane parallel to $l$ and $m$ respectively, and $w_{ij}$ is the baseline coordinate along the line of sight. 

In practice, the visibilities are affected by predominantly antenna-based complex gain factors which may vary with time, frequency, viewing direction and antenna location. Therefore, the observed visibility is defined as
\begin{equation}
V_{ij,obs} = \int \int a_i(l,m) a_j^{\dagger}(l,m) I(l,m) e^{-2\pi i \left[u_{ij}l + v_{ij}m + w_{ij}(n-1) \right]} \frac{dl dm}{n}
\end{equation}
where $^\dagger$ denotes a complex conjugate. The process of calibration tunes these antenna gains ($a_i(l,m)$ and $ a_j^{\dagger}(l,m) $) such that the calibrated or antenna gain corrected visibilities best match the sky model, ie. $V_{ij,obs} \approx V_{ij}^{\rm{sky model}}$. The sky model gives a coarse model of the (brightest) radio sources in the sky. \\


One can rewrite the RIME (Eqn. \ref{eqn:RIME}) as a linear combination of all individual sources $k$. To this end, we approximate the radio sky by a discrete number of isolated, invariant sources of finite angular extent. 
\begin{equation}
    V_{ij} = \sum_k V_{ij,k} = \sum_k \int \int I_k(l,m) e^{-2\pi i \left[u_{ij}l + v_{ij}m + w_{ij}(n-1) \right]} \frac{dl dm}{n}
\end{equation}

The `subtracted' visibilities refer to the visibilities where we have subtracted the sky model for each source. Mathematically this can be expressed as

\begin{equation}
    V_{ij, sub} = V_{ij,obs} - \sum_k \left(g_{ik} g^{\dagger}_{jk} \right)^{-1}V_{ij,k}^{\rm{model}} 
\end{equation}
with $g_{ik} = g_i(l_k, m_k) \approx a_{ik}^{-1}$, the inverse of the antenna gains. In case of a perfect calibration, this subtracted visibility is zero.\\

\section{Point spread function artefacts} \label{app:psf_artefacts}

\begin{figure*}
    \includegraphics[width=\textwidth]{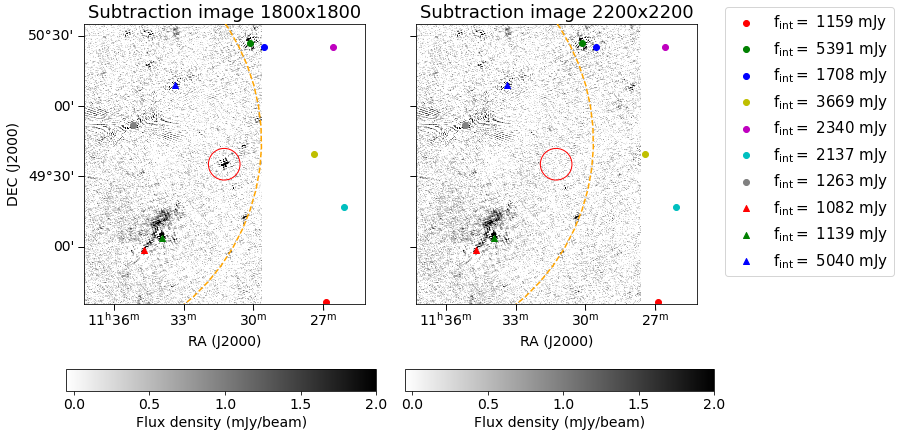}
    \caption{Example of a 'fake' transient source that is likely introduced by a bright source just at the boundary of the image. The left panel shows the a subtracted image of 1800 by 1800 pixels, while the right image shows a subtracted image of the same sky area, imaged with the same imaging settings except for increasing the image size to 2200 by 2200 pixels. The red circle indicates the location of the fake transient source present in the left image. The dashed circle show the $1.5 ^\circ$ radius that we use for filtering. The colored dots indicate the locations bright sources in the LoTSS source catalogue.}
    \label{fig:psf_artefact}
\end{figure*}

In the methods section we indicated the necessity to image 2200 by 2200 pixels, even though we are only using sources from the inner 1800 by 1800 pixels. Figure \ref{fig:psf_artefact} shows an example of a fake transient source that is likely introduced by a bright source just at the boundary of the image. The left panel shows the a subtracted image of 1800 by 1800 pixels, while the right image shows a subtracted image of the same sky area, imaged with the same imaging settings except for increasing the image size to 2200 by 2200 pixels. The red circle indicates the location of the fake transient source present in the left image. The dashed circle show the $1.5 ^\circ$ radius that we use for filtering. The colored dots indicate the locations bright sources in the LoTSS source catalogue. We think that for example the 1.7 Jy source (indicated by the navy blue dot) just beyond the 1800 by 1800 limit in the left image could introduce point spread function (PSF) artefacts quite far down the rest of the image. Simply increasing the image size a bit shows that the bright blob in the red circle in the left image is not a real transient candidate.

\begin{figure}
    \centering
    \includegraphics[width=\linewidth]{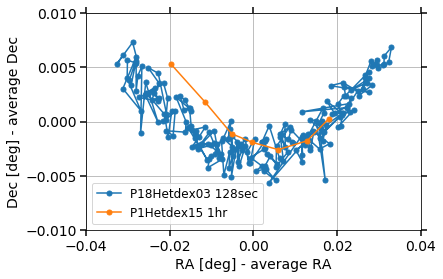}
    \caption{Example trajectory of two transient candidates, that were excluded after reimaging on a 2200 by 2200 pixel image. The right ascension and declination minus the average location of the source are shown for a source found in 2 minute subtracted image snapshots in blue, and a source found in 1 hour snapshots, shown in orange. } \label{fig:example_trajectory_psf_artefact}
\end{figure}

Initially, it was unclear what the origin of these transient candidates was. There was a strong suspicion that these types of sources were artefacts, because the source location seemed to move throughout the observation. In Figure \ref{fig:example_trajectory_psf_artefact} we show the trajectory of two transient candidates, that were excluded after reimaging on a 2200 by 2200 pixel image. The figure shows the right ascension and declination minus the average location of the source. It is clear that for both the example shown in blue, found in 2 minute subtracted image snapshots and the example shown in orange, found in 1 hour snapshots, the source seems to follow a particular trajectory.

\section{Transient candidates after visual inspection} \label{app:all_candidates}
\begin{figure}
    \centering
    \includegraphics[width=\linewidth]{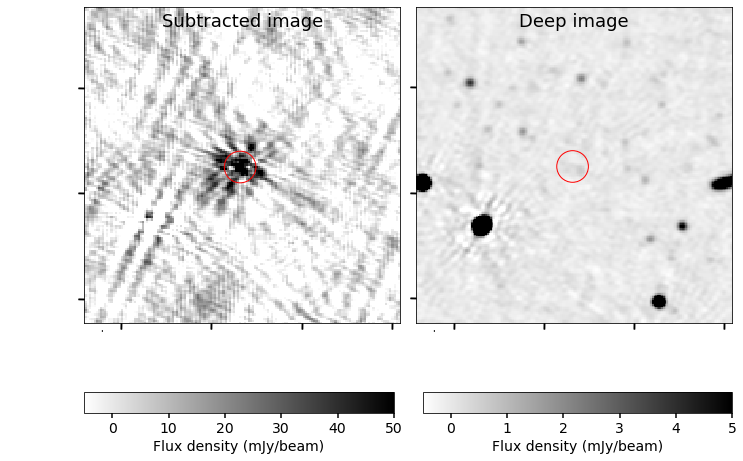}
    \caption{Transient candidate identified in three consecutive 8 second snapshots. \label{fig:appendix_candidate_P_8sec}}
\end{figure}

\begin{figure}
    \centering
    \includegraphics[width=\linewidth]{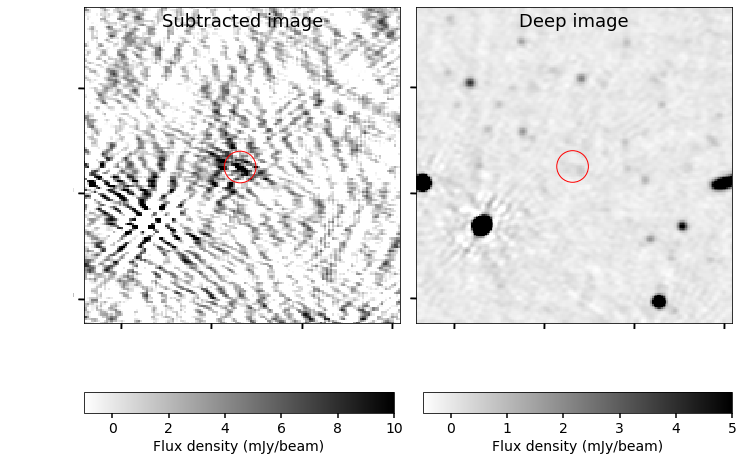}
    \caption{Transient candidate identified in one 2 minute snapshot. \label{fig:appendix_candidate_P_128sec}}
\end{figure}

\begin{figure}
    \centering
    \includegraphics[width=\linewidth]{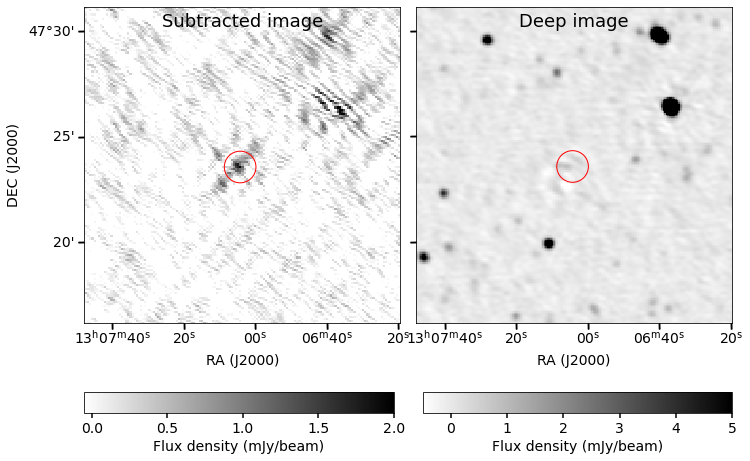}
    \caption{Transient candidate in the P35Hetdex10 field identified in all 1 hour snapshots. \label{fig:appendix_candidate_P35Hetdex10}}
\end{figure}

The figures in this Appendix show all transient candidate sources that are left after a first round of visual inspection. These sources did not immediately fall within one of the two categories of sources that are normally vetted by visual inspection (subtraction artefacts of bright sources, or sources that are associated with a faint deep fields source). The transient candidate shown in Figure \ref{fig:appendix_candidate_P_8sec} and Figure \ref{fig:appendix_candidate_P_128sec} is discussed in detail in Section \ref{sec:results_candidate1}. The transient candidate shown in Figure \ref{fig:appendix_candidate_P35Hetdex10} is discussed in more detail in Section \ref{sec:results_candidate2}. \\

\begin{figure}
    \centering
    \includegraphics[width=\linewidth]{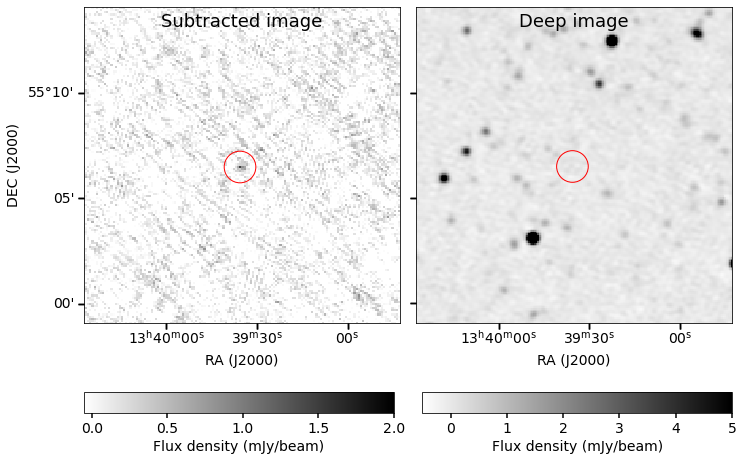}
    \caption{Transient candidate in the P205+55 field identified in one 1 hour snapshot with an snr of 6.6.}
    \label{fig:appendix_candidate_205+55}
\end{figure}

\begin{figure}
    \centering
    \includegraphics[width=\linewidth]{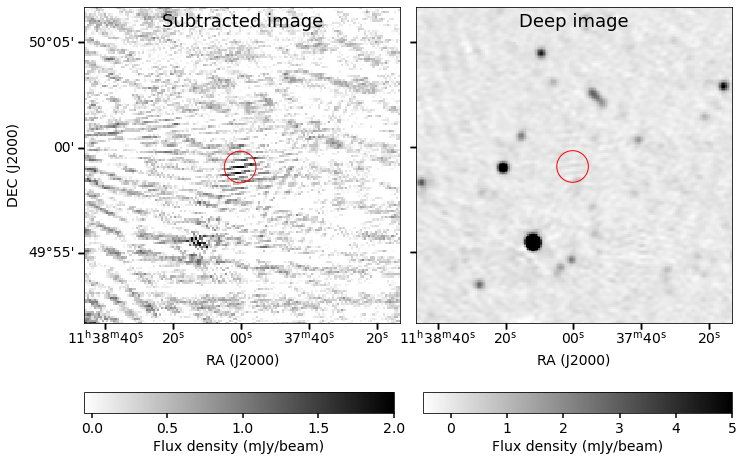}
    \caption{Transient candidate in the P14Hetdex04 field identified in the first and second 1 hour snapshot ith an snr of 5.5 and 6.7 respectively.}\label{fig:appendix_candidate_P14Hetdex04}
\end{figure}

\begin{figure}
    \centering
    \includegraphics[width=\linewidth]{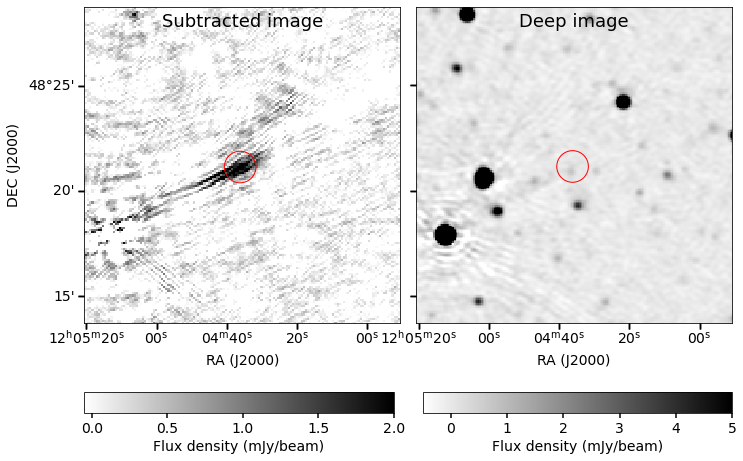}
    \caption{Transient candidate in the P19Hetdex17 field identified in one 1 hour snapshot with an snr of 6.1.}
    \label{fig:appendix_candidate_P19Hetdex17}
\end{figure}

\begin{figure}
    \centering
    \includegraphics[width=\linewidth]{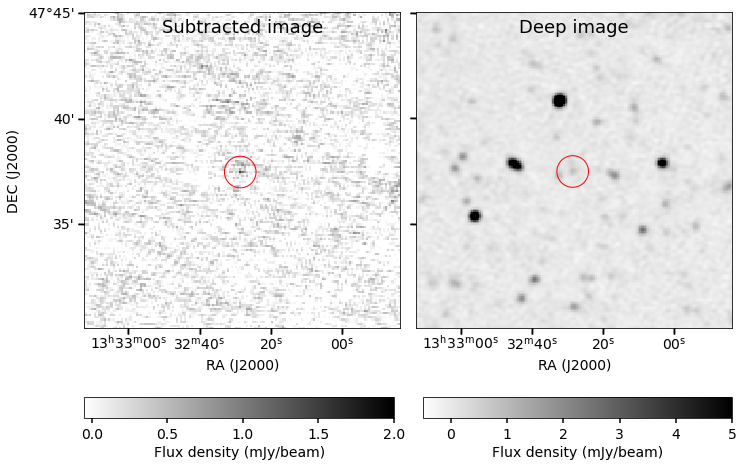}
    \caption{Transient candidate in the P39Hetdex19 field identified in the third 1 hour snapshot with an snr of 6.6.}
    \label{fig:appendix_candidate_P39Hetdex19}
\end{figure}

\begin{figure}
    \centering
    \includegraphics[width=\linewidth]{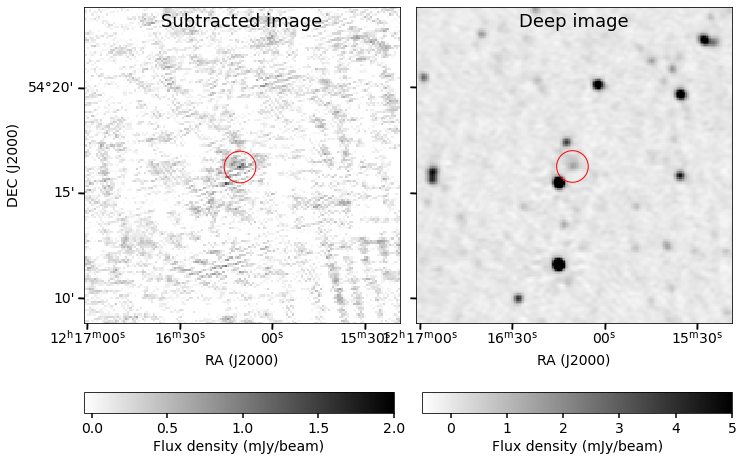}
    \caption{Transient candidate in the P182+55 field identified in the third 1 hour snapshot with an snr of 6.1.}
\label{fig:appendix_candidate_P182+55}
\end{figure}

\begin{figure}
    \centering
    \includegraphics[width=\linewidth]{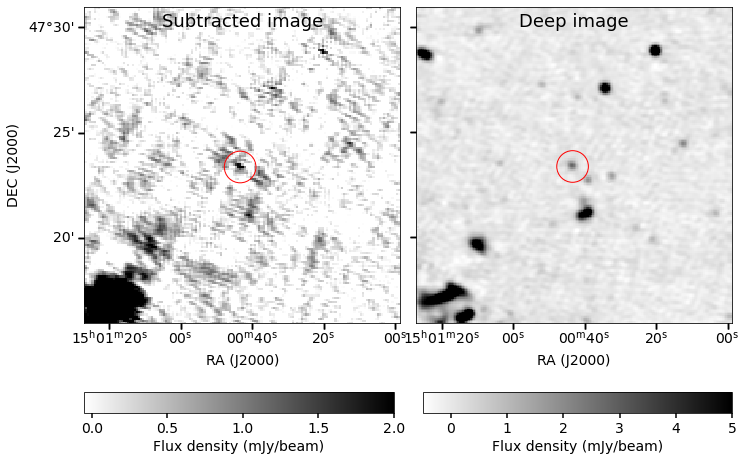}
    \caption{Transient candidate in the P225+47 field identified in the fourth and seventh 1 hour snapshot with an snr of 5.5 and 6.4 respectively.}
\label{fig:appendix_candidate_P225+47_0}
\end{figure}

\begin{figure}
    \centering
    \includegraphics[width=\linewidth]{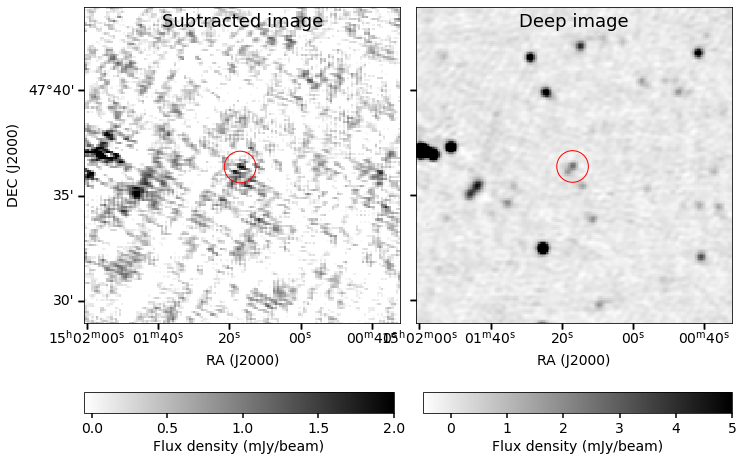}
    \caption{Transient candidate in the P225+47 field identified in the sixth and seventh 1 hour snapshot with an snr of 5.1 and 6.2 respectively.}
\label{fig:appendix_candidate_P225+47_1}
\end{figure}

\begin{figure}
    \centering
    \includegraphics[width=\linewidth]{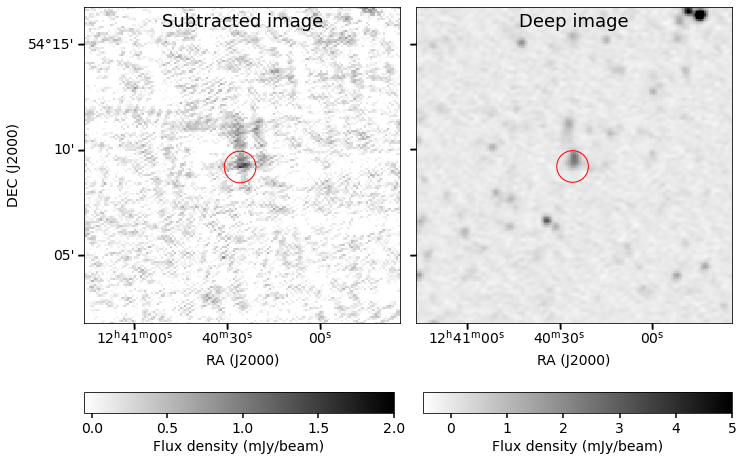}
    \caption{Transient candidate in the P191+55 field identified in the third  1 hour snapshot with an snr of 6.3.}
\label{fig:appendix_candidate_P191+55}
\end{figure}

We decide to not follow up on the candidates presented in Figure \ref{fig:appendix_candidate_205+55}, \ref{fig:appendix_candidate_P14Hetdex04} and  \ref{fig:appendix_candidate_P19Hetdex17} because the source in the subtracted image is not dirty-beam shaped, unlike expected (see Section \ref{sec:simulations}). Finally, the candidates shown in Figures \ref{fig:appendix_candidate_P39Hetdex19} to \ref{fig:appendix_candidate_P191+55} are disregarded because upon closer inspection they are associated with faint sources in the deep field.



\bsp	
\label{lastpage}
\end{document}